\documentclass[fleqn,usenatbib]{mnras}
\usepackage{newtxtext,newtxmath}
\usepackage[T1]{fontenc}
\usepackage{ae,aecompl}

\usepackage{graphicx}	
\usepackage{amsmath}	
\usepackage{amssymb}	

\newcommand{\rmaxs}{\ifmmode{R_{\rm{\sigma}}^{\rm{max}}}\else{$R_{\rm{\sigma}}^{\rm{max}}$}\fi}
\newcommand{\recirc}{\ifmmode{R_{\rm{e,c}}}\else{$R_{\rm{e,c}}$}\fi}
\newcommand{\re}{\ifmmode{R_{\rm{e}}}\else{$R_{\rm{e}}$}\fi}
\newcommand{\ret}{\ifmmode{R_{\rm{e/2}}}\else{$R_{\rm{e/2}}$}\fi}
\newcommand{\retwo}{\ifmmode{R_{\rm{e/2}}}\else{$2R_{\rm{e}}$}\fi}
\newcommand{\aee}{\ifmmode{a_{\rm{e}}}\else{$a_{\rm{e}}$}\fi}
\newcommand{\kpa}{\ifmmode{PA_{\rm{kin}}}\else{$PA_{\rm{kin}}$}\fi}
\newcommand{\ee}{\ifmmode{\epsilon_{\rm{e}}}\else{$\epsilon_{\rm{e}}$}\fi}
\newcommand{\epsintr}{\ifmmode{\epsilon_{\rm{intr}}}\else{$\epsilon_{\rm{intr}}$}\fi}

\newcommand{\lr}{\ifmmode{\lambda_R}\else{$\lambda_{R}$}\fi}
\newcommand{\lre}{\ifmmode{\lambda_{R_{\rm{e}}}}\else{$\lambda_{R_{\rm{e}}}$}\fi}
\newcommand{\lret}{\ifmmode{\lambda_{R_{\rm{e/2}}}}\else{$\lambda_{R_{\rm{e/2}}}$}\fi}
\newcommand{\lretwo}{\ifmmode{\lambda_{2R_{\rm{e}}}}\else{$\lambda_{2R_{\rm{e}}}$}\fi}

\newcommand{\flr}{\ifmmode{f_{\lambda_{R}}}\else{$f_{\lambda_{R_{\rm{e}}}}$}\fi}
\newcommand{\flre}{\ifmmode{f_{\lambda_{R_{\rm{e}}}}}\else{$f_{\lambda_{R_{\rm{e}}}}$}\fi}
\newcommand{\flret}{\ifmmode{f_{\lambda_{R_{\rm{e/2}}}}}\else{$f_{\lambda_{R_{\rm{e/2}}}}$}\fi}

\newcommand{\vs}{\ifmmode{V / \sigma}\else{$V / \sigma$}\fi}
\newcommand{\vse}{\ifmmode{(V / \sigma)_{\rm{e}}}\else{$(V / \sigma)_{\rm{e}}$}\fi}
\newcommand{\vset}{\ifmmode{(V / \sigma)_{\rm{e/2}}}\else{$(V / \sigma)_{\rm{e/2}}$}\fi}
\newcommand{\vsetwo}{\ifmmode{(V / \sigma)_{\rm{2e}}}\else{$(V / \sigma)_{\rm{2e}}$}\fi}

\newcommand{\vobs}{\ifmmode{V_{\rm{obs}}}\else{$V_{\rm{obs}}$}\fi}
\newcommand{\sobs}{\ifmmode{\sigma_{\rm{obs}}}\else{$\sigma_{\rm{obs}}$}\fi}

\newcommand{\se}{\ifmmode{\sigma_{\rm{e}}}\else{$\sigma_{\rm{e}}$}\fi}
\newcommand{\vrms}{\ifmmode{\langle v^2_{\rm{rms}}\rangle}\else{$\langle v^2_{\rm{rms}}\rangle$}\fi}
\newcommand{\vrmse}{\ifmmode{\langle v^2_{\rm{rms}}\rangle_{\rm{e}}}\else{$\langle v^2_{\rm{rms}}\rangle_{\rm{e}}$}\fi}
\newcommand{\sqvrmse}{\ifmmode{\langle v^2_{\rm{rms}}\rangle^{1/2}_{\rm{e}}}\else{$\langle v^2_{\rm{rms}}\rangle^{1/2}_{\rm{e}}$}\fi}

\newcommand{\kms}{\ifmmode{\,\rm{km}\, \rm{s}^{-1}}\else{$\,$km$\,$s$^{-1}$}\fi}
\newcommand{\msun}{\ifmmode{~\rm{M}_{\odot}}\else{M$_{\odot}$}\fi}
\newcommand{\mstar}{\ifmmode{M_{\star}}\else{$M_{\star}$}\fi}
\newcommand{\logm}{\ifmmode{\log(M_{\star}/M_{\odot})}\else{$\log(M_{\star}/M_{\odot})$}\fi}
\newcommand{\mm}{\ifmmode{M_{\star}/M_{\odot}}\else{$M_{\star}/M_{\odot}$}\fi}

\newcommand{\ser}{S\'ersic}

\newcommand{\ea}{\textsc{eagle}}
\newcommand{\eap}{\textsc{eagle$^+$}}
\newcommand{\hy}{\textsc{hydrangea}}
\newcommand{\ha}{\textsc{horizon-agn}}
\newcommand{\ill}{\textsc{illustris}}
\newcommand{\ma}{\textsc{magneticum}}

\newcommand{\at}{\ifmmode{\rm{ATLAS}^{\rm{3D}}}\else{ATLAS$^{\rm{3D}}$}\fi}

\title[Comparing IFS Galaxy Observations with Simulations]{The SAMI Galaxy Survey: comparing 3D spectroscopic observations with 
galaxies from cosmological hydrodynamical simulations}

\author[Jesse van de Sande]{Jesse van de Sande$^{1,2}\thanks{jesse.vandesande@sydney.edu.au},$
Claudia D.P. Lagos$^{2,3}, $
Charlotte Welker$^{2,3}, $
\newauthor
Joss Bland-Hawthorn$^{1,2}, $
Felix Schulze$^{4,5}, $
Rhea-Silvia Remus$^{4}, $
Yannick Bah\'e$^{\,6}, $
\newauthor
Sarah Brough$^{2,7}, $
Julia J. Bryant$^{1,2,8}, $
Luca Cortese$^{2,3}, $
Scott M. Croom$^{1,2}, $
\newauthor
Julien Devriendt$^{9}, $
Yohan Dubois$^{10}, $
Michael Goodwin$^{11}, $
Iraklis S. Konstantopoulos$^{12}, $
\newauthor
Jon S. Lawrence$^{11}, $
Anne M. Medling$^{13,14,15}, $
Christophe Pichon$^{10,16}, $
\newauthor
Samuel N. Richards$^{17}, $
Sebastian F. Sanchez$^{18}, $
Nicholas Scott$^{1,2} $ 
and Sarah M. Sweet$^{2,19}$
\\
\\
Affiliations are listed at the end of the paper
}

\date{Accepted XXX. Received YYY; in original form ZZZ}

\pubyear{2019}

\begin{document}
\label{firstpage}
\pagerange{\pageref{firstpage}--\pageref{lastpage}}
\maketitle



\begin{abstract}
Cosmological hydrodynamical simulations are rich tools to understand the build-up of stellar mass and angular momentum in galaxies, but require some level of calibration to observations. We compare predictions at $z\sim0$ from the \textsc{eagle}, \textsc{hydrangea}, \textsc{horizon-agn}, and \textsc{magneticum} simulations with integral field spectroscopic (IFS) data from the SAMI Galaxy Survey, ATLAS$^{\rm{3D}}$, CALIFA and MASSIVE surveys. The main goal of this work is to  simultaneously compare structural, dynamical, and stellar population measurements in order to identify key areas of success and tension. We have taken great care to ensure that our simulated measurement methods match the observational methods as closely as possible, and we construct samples that match the observed stellar mass distribution for the combined IFS sample. We find that the \textsc{eagle} and \textsc{hydrangea} simulations reproduce many galaxy relations but with some offsets at high stellar masses. There are moderate mismatches in $R_{\rm{e}}$ (+), $\epsilon$ ($-$), $\sigma_{\rm{e}}$ ($-$), and mean stellar age ($+$), where a plus sign indicates that quantities are too high on average, and minus sign too low. The \textsc{horizon-agn} simulations qualitatively reproduce several galaxy relations, but there are a number of properties where we find a quantitative offset to observations. Massive galaxies are better matched to observations than galaxies at low and intermediate masses. Overall, we find mismatches in $R_{\rm{e}}$ (+), $\epsilon$ ($-$), $\sigma_{\rm{e}}$ ($-$) and $(V/\sigma)_{\rm{e}}$ ($-$). \textsc{magneticum} matches observations well: this is the only simulation where we find ellipticities typical for disk galaxies, but there are moderate differences in $\sigma_{\rm{e}}$ ($-$), $(V/\sigma)_{\rm{e}}$ ($-$) and mean stellar age (+). Our comparison between simulations and observational data has highlighted several areas for improvement, such as the need for improved modelling resulting in a better vertical disk structure, yet our results demonstrate the vast improvement of cosmological simulations in recent years.
\end{abstract}



\section{Introduction}
\label{sec:introduction}

In the present-day Universe, the majority of galaxies ($>85$ percent) are consistent with being axisymmetric rotating oblate spheroids and only a minor fraction of galaxies have complex dynamics \citep[for a review, see][]{cappellari2016}. The ratio of ordered to random stellar motion in galaxies has a strong dependence on luminosity or stellar mass \citep{illingworth1977,davies1983,emsellem2011,brough2017,veale2017b,vandesande2017b,green2018}, which suggests a link between the build-up of stellar mass and angular momentum over time. Many theoretical studies are aimed at explaining the build-up and removal of angular momentum in galaxies through mergers \citep[][and citations within]{naab2014}.

Binary galaxy merger simulations are a commonly used tool for studying the dynamical evolution of galaxies. These simulations showed that most merger remnants are consistent with being fast rotating galaxies \citep{bois2010,bois2011}, similar to what is found in the observational data. The dominant process for creating realistic slow rotating galaxies, however, is still a matter of debate \citep[e.g.,][]{bendo2000,jesseit2009,bois2011}. The mass ratio of the progenitors in binary-disk mergers appears to be the most critical parameter for creating slow rotators, but there is also a strong dependence on specific spin-orbit alignments \citep{jesseit2009,bois2010,bois2011}. Merger remnants formed from dissipational (wet) mergers of equal-mass disk galaxies better match the observed data than dissipationless (dry) merger remnants \citep{cox2006}. This suggests that the presence of gas during mergers is crucial for creating slow rotators. However, this is in contrast with \citet{taranu2013} who show that dissipation is not a prerequisite for producing slow-rotating galaxies. Instead, multiple, mostly dry, minor mergers are sufficient.

To disentangle the relative importance of major and minor mergers, and large-scale environment, in changing the angular momentum in galaxies, one requires a large ensemble of simulated galaxies with a range of initial conditions. Large cosmological hydrodynamical simulations are well suited for this. These simulations follow the growth and evolution of the galaxy from high-redshift ($z\sim50$) to the present-day ($z=0$) and provide more realistic insights into the formation paths and rotational properties of galaxies as compared to previous idealized, binary merger simulations.

The success of such an approach has already been demonstrated by \citet{naab2014}, \citet{welker2017}, \citet{remus2017}, \citet{penoyre2017}, \citet{choi2017}, \citet{choi2018}, \citet{lagos2018a,lagos2018b}, and \citet{martin2018}. \citet{naab2014} use cosmological hydrodynamical zoom-in simulations of 44 individual central galaxies, and link the assembly history of these galaxies to their stellar dynamical features. Their analysis of the stellar kinematic data is done in an identical way to the \at\ kinematic observations \citep{cappellari2011a}. They find a good qualitative agreement between the simulated and observed kinematic measurements. By following the merger histories of galaxies, 
\citet{naab2014} show that there are multiple formation paths for fast and slow rotating galaxies, emphasizing the importance of studying large ensembles of simulated galaxies. 

\citet{penoyre2017} use the \ill\ simulations to follow the dynamical evolution of thousands of galaxies. They show that after $z=1$, the merger and star-formation histories of slow and fast rotator progenitors start to differ. In contrast to \cite{naab2014} and \citet{lagos2018b}, they find no major difference between the effects of gas-rich and gas-poor mergers. Minor mergers also appear to have little correlation with the spin of galaxies. \citet{lagos2018a} use \ea\ \citep{schaye2015,crain2015} to analyse the effect galaxy mergers (with different parameters) have on the specific angular momentum ($j_{\star}$) of galaxies. They show that, on average, dry mergers reduce $j_{\star}$ by $\approx30$ per cent, while wet mergers increase $j_{\star}$ by $\approx10$ per cent. \citet{choi2018} and \citet{lagos2018b} focus on the impact of galaxy mass and environment on the spin-down of galaxies, using the \ha\ \citep{dubois2014} and \ea\ simulations respectively. Both studies agree with the observational results from \cite{veale2017a}, \citet{brough2017}, and \citet{green2017} that galaxy stellar mass plays a more dominant role in changing the spin-parameter proxy (\lr) of galaxies than environment. For satellite early-type galaxies, non-merger-induced tidal perturbations also appear to play a bigger role than mergers in lowering the galaxy spin parameter \citep{choi2018}.
 
These specific angular momentum and spin parameter evolution predictions, however, assume that other galaxy parameters and scaling relations at $z\sim0$ are also well-matched to observations. Most simulated galaxy populations appear to overlap in terms of their dynamics and shapes (e.g., \citealt{penoyre2017} using \ill; \citealt{lagos2018b} using \ea\ and \hy), but some mismatch between the observations and simulations is also present \citep[][using \ha]{choi2018}. The validity of simulation predictions become doubtful if the main parameter that is being used to make the predictions matches well with observations, while other parameters do not. Thus, a detailed comparison between multiple observational properties of galaxies from observations and simulations is needed to support the idea that conclusion from simulations apply to the real Universe.

Integral field spectroscopic (IFS) observations are ideally suited for a comparison with simulations. IFS galaxy surveys provide a unique opportunity to compare resolved two-dimensional stellar dynamical measurements across a large range of galaxy stellar masses and morphologies. Furthermore, IFS samples are typically selected from larger surveys that contain a wealth of ancillary data including structural parameters, stellar masses, and large scale environmental estimates. Cosmological hydrodynamical simulations are now also capable of creating large samples of mock-galaxies with dynamical observations with high enough spatial resolution to resolve some of the inner dynamical structures of galaxies.

In this paper, we compare structural, resolved dynamical, and stellar population observations of mock galaxies from the \ea, \hy\ \citep{bahe2017}, \ha, and \ma\ simulations to observations from the Sydney-AAO Multi-object Integral field spectrograph (SAMI) Galaxy Survey \citep{croom2012, bryant2015}, the \at\ Survey \citep{cappellari2011a}, the CALIFA survey \citep{sanchez2012}, and the MASSIVE survey \citep{ma2014}. The paper is organized as follows: Section \ref{sec:obs_data} and \ref{sec:sim_data} respectively present the data from the observations and simulations. In Section \ref{sec:comp_obs_sim} we compare several observational relationships between stellar mass, size and dynamical parameters with the predictions from the simulations. We review previous comparison studies in Section \ref{sec:previous}. The implications of these matches and mismatches are discussed and summarised in the Section \ref{sec:summary_discussion}. 

Throughout the paper we assume a $\Lambda$CDM cosmology with $\Omega_\mathrm{m}$=0.3, $\Omega_{\Lambda}=0.7$, and $H_{0}=70$ km s$^{-1}$ Mpc$^{-1}$. Furthermore, we adopt a \citet{chabrier2003} stellar initial mass function (IMF).

\section{Observational Data}
\label{sec:obs_data}

\subsection{SAMI Galaxy Survey}
SAMI is a multi-object IFS mounted at the prime focus of the 3.9m Anglo Australian Telescope (AAT). It employs 13 revolutionary imaging fibre bundles, or \textit{hexabundles} \citep{blandhawthorn2011,bryant2011,bryant2012a,bryant2014} that are manufactured from 61 individual fibres with 1\farcs6 angle on sky. Each hexabundle covers a $\sim15^{\prime\prime}$ diameter region on the sky, has a maximal filling factor of 75\%, and is deployable over a 1$^\circ$ diameter field of view. All 819 fibres, including 26 individual sky fibres, are fed into the AAOmega dual-beamed spectrograph \citep{saunders2004, smith2004, sharp2006}. 

The SAMI Galaxy Survey \citep{croom2012, bryant2015} has finished observations, targeting over 3000 galaxies covering a broad range in galaxy stellar mass (M$_* = 10^{8}-10^{12}$\msun) and galaxy environment (field, groups, and clusters) between redshift $0.004<z<0.095$. Here we use internal data release v0.10.1 that contains 2528 galaxies. SAMI's angular fibre size results in spatial resolutions of 1.6 kpc per fibre at $z=0.05$. Field and group targets were selected from four volume-limited galaxy samples derived from cuts in stellar mass in the Galaxy and Mass Assembly Survey (GAMA) G09, G12 and G15 regions \citep{driver2011}. GAMA is a major campaign that combines a large spectroscopic survey of $\sim$300,000 galaxies carried out using the AAOmega multi-object spectrograph on the AAT, with a large multi-wavelength photometric data set. SAMI Galaxy Survey cluster targets were obtained from eight high-density cluster regions sampled within radius $R_{200}$ with the same stellar mass limit as for the GAMA fields \citep{owers2017}. 

For the SAMI Galaxy Survey, the 580V and 1000R grating are used in the blue (3750-5750\AA) and red (6300-7400\AA) arm of the spectrograph, respectively. This results in a resolution of R$_{\rm{blue}}\sim 1810$ at 4800\AA, and R$_{\rm{red}}\sim4260$ at 6850\AA\ \citep{vandesande2017}. In order to create data cubes with 0\farcs5 spaxel size, all observations are carried out using a six to seven position dither pattern \citep{sharp2015,allen2015}. 

All reduced data-cubes and stellar kinematic data products in the GAMA fields are available on: \url{https://datacentral.org.au/}, as part of the first and second SAMI Galaxy Survey data release \citep{green2017,scott2018}.

\subsubsection{Ancillary Data}

For galaxies in the GAMA fields, we use the aperture matched $g$ and $i$ photometry from the GAMA catalogue  \citep{hill2011,liske2015}, measured from reprocessed SDSS Data Release Seven \citep{york2000, kelvin2012}, to derive $g-i$ colours. For the cluster environment, photometry from the SDSS \citep{york2000} and VLT Survey Telescope ATLAS imaging data are used \citep{shanks2013,owers2017}. From the rest-frame i-band absolute magnitude and $g-i$ color, stellar masses are derived by using the color-mass relation as outlined in \citet{taylor2011}. A \citet{chabrier2003} stellar IMF and exponentially declining star formation histories are assumed in deriving the stellar masses. For more details see \citet{bryant2015}.

Effective radii, ellipticities, and positions angles are derived using the Multi-Gaussian Expansion \citep[MGE;][]{emsellem1994,cappellari2002} technique and the code from \citet{scott2013} on imaging from the GAMA-SDSS \citep{driver2011}, SDSS \citep{york2000}, and VST \citep{shanks2013,owers2017}. 
We define \re\ as the semi-major axis effective radius, and $\recirc = \re\times\sqrt{1-\epsilon}$ as the circularised effective radius. The ellipticity of the galaxy within one effective radius is defined as $\epsilon_{\rm{e}}$, measured from the best-fitting MGE model. For more details, we refer to D'Eugenio et al. (in prep). 

We use visual morphological classifications based on the Hubble type \citep{hubble1926} following the scheme used by \citet{kelvin2014}. The classifications are determined from SDSS DR9 and VST {\it gri} colour images. Late- and early-types are divided according to their shape, presence of spiral arms and/or signs of star formation. Pure bulges are then classified as ellipticals (E) and early-types with disks as S0s. Similarly, late-types with only a disk component are classified as late-spirals, while disk plus bulge late types are early-spirals \citep[for more details see][]{cortese2016}.

\subsubsection{Stellar Kinematics}
\label{subsubsec:stelkin_sami}

The stellar kinematic measurements for the SAMI Galaxy Survey are described in detail in \citet{vandesande2017}. In summary, we use the penalized pixel fitting code \citep[pPXF;][]{cappellari2004,cappellari2017} assuming a Gaussian line of sight velocity distribution (LOSVD). The red arm spectral resolution is convolved to match the instrumental resolution in the blue. Both blue and red are then rebinned and combined onto a logarithmic wavelength scale with constant velocity spacing (57.9 \kms). MILES stellar library \citep{sanchezblazquez2006,falconbarroso2011} spectra are used for deriving a set of radially varying optimal templates using the SAMI annular binned spectra. For each individual spaxel, \textsc{pPXF} is allowed to use the optimal templates from the annular bin in which the spaxel is located as well as the optimal templates from neighbouring annular bins. The uncertainties on the LOSVD parameters are estimated from 150 simulated spectra.

We use the quality criteria for SAMI Galaxy Survey data as described in \citet{vandesande2017}: signal-to-noise (S/N) $>3$\AA$^{-1}$, the observed velocity dispersion \sobs $>$ FWHM$_{\rm{instr}}/2 \sim 35$\kms\ where the FWHM is the full-width at half-maximum of the instrumental resolution or line-spread function, $V_{\rm{error}}<30$\kms \citep[Q$_1$ from][]{vandesande2017}, and $\sigma_{\rm{error}} < \sobs *0.1 + 25$\kms\ (Q$_2$)

We visually inspect all 2528 SAMI kinematic maps, and flag and exclude 87 galaxies with irregular kinematic maps due to nearby objects or mergers that influence the stellar kinematics of the main object. Another 533 galaxies are excluded where the radius out to which we can accurately measure the stellar kinematics or \re\ is less than the half-width at half-maximum of the PSF (HWHM$_{\rm{PSF}}$). Furthermore, we set the observational mass limit for stellar kinematic measurement at $\mstar=5\times10^9\msun$ or $\logm=9.7$, similar to the simulations, which excludes another 332 galaxies. Finally, throughout the paper, we only use galaxies when where we can accurately derive \vs\ out to one \re\ (see Section \ref{subsec:mass_vsigma}). This brings the final number of galaxies from the SAMI Galaxy Survey to 1558.

We note that while the total number of targeted galaxies in the GAMA fields is significantly higher as compared to cluster targets, due to higher stellar mass limit and lower redshift range of the cluster sample, the stellar kinematic success rate in clusters is significantly higher. Hence, we end up with a relatively large fraction of cluster galaxies in the final SAMI stellar kinematic sample ($\sim47$ percent).

\subsubsection{Stellar Population Age}
\label{subsubsec:stelpop_sami}

We derive luminosity-weighted stellar population ages using 11 Lick indices in the SAMI blue spectral range following the method outlined in \citet{scott2017}. Lick indices are converted into single stellar population (SSP) equivalent age using stellar population synthesis models \citep{schiavon2007} that predict Lick indices as a function of $\log \rm{Age}$, metallicity [Z/H], and [$\alpha$/Fe]. We then determine the SSP that best reproduces the measured Lick indices using a $\chi^2$ minimisation approach. Typical uncertainties are $\pm0.15$ dex in $\log$ Age. Because the stellar population parameters are derived from using individual Lick indices rather than full spectral fitting, our results are relatively insensitive to dust.

\subsection{\at\ Survey}
\label{subsec:at_data}

We use a combined sample of 260 early-type galaxies from the SAURON survey \citep{dezeeuw2002} and \at\ Survey \citep{cappellari2011a} that were observed with the SAURON spectrograph \citep{bacon2001}. The SAURON survey adopted an instrumental spectral resolution of 4.2\AA\ FWHM ($\sigma_{\rm{instr}}$ = 105 \kms) and cover the wavelength range of 4800-5380\AA. \at\ galaxies were observed with a higher resolution of 3.9\AA\ FWHM ($\sigma_{\rm{instr}}$ = 98\kms). The data were Voronoi binned \citep{cappellari2003} with a target signal-to-noise per bin of 40. The stellar kinematics were extracted using \textsc{pPXF} with stellar templates from the MILES stellar library \citep[see ][]{cappellari2011a}.

We use the publicly available unbinned flux data cubes (V1.0\footnote{http://www-astro.physics.ox.ac.uk/atlas3d/}) and the 2D Voronoi binned stellar kinematic maps \citep{emsellem2004,cappellari2011a}. We exclude galaxy NGC 0936 from the sample because no unbinned flux data is available. Circularised size measurements are taken from \citet{cappellari2011a}, corrected to semi-major axis effective radii using the global ellipticities from \cite{krajnovic2011} ($\re=\recirc/\sqrt{1-\epsilon}$). We use ellipticities at one effective radius from \citet{emsellem2011}, and position angles from \cite{krajnovic2011}. Stellar masses are calculated from the $r$-band luminosity and mass-to-light ratio as presented in \citet{cappellari2013a,cappellari2013b}, converted to a \citet{chabrier2003} IMF by subtraction a constant factor of 0.24 dex \citep[see for example][]{muzzin2009}. Visual morphologies (T-type) are from \citet{cappellari2011a}. We obtained luminosity-weighted stellar population ages from Table 3 of \citet{mcdermid2015} that are based on Lick indices measurements and single stellar population models from \citet{schiavon2007}. As described in Appendix \ref{sec:app_age_sami_a3d}, we subtract 0.23 dex from all \at\ stellar ages to correct a median offset from the SAMI sample. Above a stellar mass limit of $\logm=9.7$, the \at\ sample contains 240 galaxies.

\subsection{CALIFA Survey}
\label{subsec:califa_data}

We use a sample of 294 CALIFA galaxies with a wide range in morphology and kinematic properties from \cite{falconbarroso2017}. The CALIFA IFS data were observed with the PMAS instrument \citep{roth2005}, a $74'' \times 64''$ hexagonal fibre spectrograph, mounted at the 3.5m telescope of the Calar Alto observatory. Similar to \at, the CALIFA data are Voronoi binned to obtain spatial bins with an approximate S/N of 20 per pixel. Stellar kinematic measurements were derived from the Voronoi binned spectra with wavelength range between 3400-4750\AA\ and spectral resolution of 2.3\AA\ \citep[V1200 grating;][]{husemann2013}. Velocities $V$ and velocity dispersions $\sigma$ are extracted with the \textsc{pPXF} code in combination with 330 stellar templates selected from the Indo-U.S. spectral library \citep{valdes2004}. The 2D kinematic maps are publicly available as part of the \mbox{CALIFA} DR3 \citep{sanchez2016}\footnote{http: //califa.caha.es}. 

We use the stellar masses, visual morphologies, semi-major axis effective radii, position angles, and ellipticities at one effective radius as presented in Table 1 from \citeauthor{falconbarroso2017} \citep[\citeyear{falconbarroso2017}; see also][]{walcher2014}. Luminosity-weighted stellar population ages are gathered from table C.2 of \citet{gonzalez2015}. These ages are estimated from full spectral fitting using the \textsc{starlight} code \citep{cidfernandes2005}, and the spectral base GMe models that are a combination of SSP spectra provided by \citet{vazdekis2010} and \citet{gonzalez2005}. Measurements within one effective radius are not available, instead we derive the average of the "central" ([0]) and "at 1 half-light-radius" mean stellar ages. Finally, we note that CALIFA galaxies are selected based on angular isophotal diameter \citep[$45'' \leq D25 \leq 80''$][]{walcher2014}; this  biases the sample towards galaxies that are more inclined and have higher ellipticities. Above a stellar mass limit of $\logm=9.7$, the CALIFA sample contains 257 galaxies.

There are six galaxies in the CALIFA survey that overlap with \at. Similar to \citet{falconbarroso2017}, we find an excellent agreement between the dynamical measurements from both surveys. However, when comparing the stellar mass estimates, we notice that 5/6 CALIFA stellar masses are on average higher by $\sim0.22$dex as compared to \at. To investigate this offset further, we check for possible mismatches in the size-stellar mass plane and velocity dispersion-stellar mass plane. We find that the full CALIFA dataset is on average higher by $\sim0.2$ dex in stellar mass as compared to GAMA, SAMI, and \at\ data in the size-mass diagram, and to SAMI and \at\ in the \se-mass diagram. This difference could be explained by the magnitude measurements that are used by the CALIFA survey as compared to SDSS. It was shown in \citet{walcher2014} that the CALIFA growth curve magnitude measurements are brighter than the magnitudes from the SDSS $petroMag_r$ ($\Delta(g-mag) = 0.34$). While other CALIFA stellar mass estimates exist \citep[e.g.,][]{gonzalez2015,sanchez2016b}, when comparing these stellar masses to the overlapping \at\ galaxies, or with dynamical mass measurements (see Section \ref{subsec:mass_mdyn}), we find that they do not provide a better match. As the main idea behind using different surveys is to create a more homogeneous data-set, we therefore decide to use the \cite{falconbarroso2017} stellar masses, but we subtract 0.2 dex to correct for the offset to GAMA, SAMI, and \at.

\subsection{MASSIVE Survey}
\label{subsec:massive_data}

The volume-limited MASSIVE IFS survey specifically targeted the $\sim100$ most massive early-type galaxies within a distance of 108 Mpc. IFS observations were done with the Mitchell fibre Spectrograph \citep{hill2008} with a $107'' \times 107''$ field of view, on the 2.7m Harlan J. Smith Telescope at McDonald Observatory. The data are spatially combined using a circular binning scheme to reach a target S/N of 20. The wavelength range is 3650-5850\AA\ and the average spectral resolution is 5\AA\ (FWHM) with a dependence on wavelength and spatial position.

\citet{veale2017a} use \textsc{pPXF} in combination with the MILES stellar library to extract six-moment kinematics, i.e., they fit for $V$, $\sigma$, and $h_3-h_6$. Here, $h_3-h_6$ are the weights of Gauss-Hermite polynomials that are used to model the deviations from a Gaussian LOSVD. Stellar kinematic maps are not publicly available; instead we use the kinematic values as presented in Table 1 of \citet{veale2018}. This will be explained in more detail in Section \ref{subsec:mass_vsigma}. Stellar velocity dispersions and ellipticities are obtained from \citet{veale2017a,veale2017b}. Ellipticities are derived from a "super-coadd" isophote, not within one effective radius \citep{ma2014}. Effective radii are from \citet{ma2014} Table 3. We use their NASA-Sloan Atlas values - which are based on SDSS imaging \citep{york2000,blanton2011} - where available, or 2MASS XSC catalogue values \citep{jarrett2000,skrutskie2006} corrected using their Eq.~4.

Following \citet{ma2014}, we use absolute $K$-band magnitudes to estimate stellar masses. However, rather than deriving stellar masses from a relation based on Jeans Anisotropic Modelling (JAM) mass-to-light ratios \citep{cappellari2013a}, we use stellar population model-based mass-to-light ratios ($\log (M/L)_{\rm{Salp}}$) from \citet{cappellari2013b}, converted to a \citet{chabrier2003} IMF:

\begin{equation}
\log_{10}(\mstar) = 10.39-0.46(M_K+23).
\end{equation}

\noindent This different choice as compared to \citet{ma2014} is motivated by the need for a homogeneous sample where all stellar masses are calculated in the same way. Above a stellar mass limit of $\logm=9.7$, the total number of MASSIVE galaxies with stellar kinematic measurements is 85. 

\subsection{Summary of Observational Data}
\label{subsec:summary_obsdata}

The combined data from the SAMI Galaxy Survey, \at, CALIFA, and MASSIVE Survey yields a total of 2140 galaxies with stellar kinematic measurements above a stellar mass of $\logm>9.7$. Approximately $\sim40$ percent of the galaxies reside in high-density cluster environments, and $\sim$66 percent are visually classified to have early-type morphology (E and S0-type).

\section{Simulation Data}
\label{sec:sim_data}

\subsection{EAGLE and HYDRANGEA Simulations}
\label{subsec:eagle_sims}

The \ea\ project \citep[Evolution and Assembly of GaLaxies and their Environments;][]{schaye2015,crain2015,mcalpine2016}, is a large set of cosmological hydrodynamic simulations that are publicly available\footnote{http://icc.dur.ac.uk/Eagle/database.php}. In this paper, we use the reference model Ref-L100N1504 with a volume of (100 Mpc)$^3$ co-moving. \ea\ and \hy\ adopt the Planck Collaboration XVI \citeyear{planck2014} cosmological parameters ($\Omega_\mathrm{m}$=0.307, $\Omega_{\Lambda}=0.693$, $H_{0}=67.77$\kms\ Mpc$^{-1}$). The dark matter particle mass is $9.7\times10^6\msun$, the initial gas particle mass is $1.81\times10^6\msun$, and the typical mass of a stellar particle is similar to the gas particle mass. The reference model was calibrated to match the $z\sim0.1$ stellar mass function and the observed relation between stellar mass and black-hole mass. The $z\sim0.1$ size-mass relation was also used as a guide to reject some stellar feedback models that led to too compact galaxies, despite reproducing the stellar mass function (see \citealt{crain2015} for details).

To provide the best global environment match to the SAMI Galaxy Survey sample, we combine \ea\ with \hy\ that consists of 24 cosmological zoom-in simulations of galaxy clusters and their environments \citep{bahe2017}. \hy\ is part of the larger Cluster-\ea\ project \citep{barnes2017}. Cluster-\ea\ is simlar to \ea\, but with different parameter values for the active galactic nuclei (AGN) feedback model, to make it more efficient. To reproduce the same ratio of field, group, and cluster galaxies as in observations, we select all \hy\ galaxies that are in groups or clusters with mass greater than $\logm_{\rm{group}}>13.85$. We note that this group mass limit for \hy\ is lower than the adopted mass for the SAMI Galaxy Survey cluster sample of $\logm_{\rm{group}}>14.25$. This lower limit was adopted to reach a fraction of $\sim40$ percent of galaxies that are in the highest-density environments, similar to observations.

We use the kinematic measurements as described in \citet{lagos2018b}, corrected to $H_{0}=70.0$\kms\ Mpc$^{-1}$ which was adopted for the observations. In summary, we extract effective radii, ellipticities, line-of-sight velocities and velocity dispersions, adopting techniques that closely match the observations. First, each galaxy is projected onto a 2D plane that is observed under two different inclinations: an edge-on view, and a random orientation (seen through the simulation $z$-axis) to mimic observations. In this 2D projection, we create a grid of pixels with size 1.5 kpc (proper), and construct an $r$-band luminosity weighted velocity distribution for each pixel. This LOSVD is fitted with a Gaussian function to estimate $V$ and $\sigma$; the rest-frame velocity is defined at the centre of the galaxy potential. 

$r$-band luminosities of star particles are derived by combining the age and metallicity of those star particles assuming a \cite{chabrier2003} IMF combined with a \cite{bruzual2003} stellar population synthesis model. The $r$-band luminosities are then obtained by convolving this model with an SDSS $r$-band bandpass. We do not include the effect of dust on the $r$-band luminosities. While dust obscuration will have a larger impact on the measurements in late-type galaxies than in early-types and depends on inclination, as the differences between mass and luminosity weighted quantities are relatively small (see Appendix \ref{sec:app_lum_mass}), we do not expect dust to change our results significantly. For a study on the impact of dust on galaxy colours in \ea, we refer to \citet{trayford2017}.

We measure the projected circularised $r$-band half-luminosity radius (\recirc) as the average from the $xy$, $xz$, and $yz$ projections. Ellipticities are calculated within one effective radius using the projected particle positions and particle luminosities using Eqs.~$1$~to~$3$ of \citet{lagos2018b}. The position angle of the major axis are defined in Eq.~4 of \citet{lagos2018b}. Stellar ages were calculated as an $r$-band luminosity weighted stellar age using all particles within $r_{50}$. 

Following \citet{lagos2018b}, we adopt a lower mass limit of $\mstar=5\times10^9\msun$, to ensure that the simulated measurements converge to better than 10 percent. Example maps of the luminosity, velocity, and velocity dispersion are shown in Fig~1 of \citet{lagos2018b}. Above the adopted stellar mass limit, the combined \ea\ and \hy\ sample contains 8,982 galaxies.

\subsection{HORIZON-AGN Simulations}
\label{subsec:horizon_agn_sims}

The cosmological hydrodynamic \ha\ simulations are described in detail in \citet{dubois2014}. In short, the simulation that we use here is run within a box with a volume of (142 Mpc)$^3$ co-moving. \ha\ adopts a cosmology compatible with the Wilkinson Microwave Anisotropy Probe 7 cosmology \citep[$\Omega_\mathrm{m}$=0.272, $\Omega_{\Lambda}=0.728$, $H_{0}=70.4$\kms\ Mpc$^{-1}$;][]{komatsu2011}. The dark matter particle mass is $8\times10^7\msun$. The hydrodynamics are computed on a grid, adaptively refining to the local density following a quasi Lagrangian scheme \citep{teyssier2002}. Cells are 1kpc wide at maximal refinement level. The adopted resolution is such that the typical mass of a stellar particle is $2\times10^6\msun$. The simulation implements the formation and evolution of black holes, and black holes can grow by gas accretion. The \ha\ project uses two modes of AGN feedback, while \ea\ adopts one. The tuning approach in \ha\ differs from \ea. Only the local black hole mass to stellar mass relation is tuned so as to ensure the predictive aspect of resulting statistical properties in the simulation.

Structural, stellar population, and stellar kinematic measurements are measured in a similar way as was done for the \ea\ simulation, corrected to $H_{0}=70.0$\kms\ Mpc$^{-1}$. We present a brief summary here, whereas a more detailed description will be presented in C. Welker et al. (in prep.). Ages and ellipticities are computed directly from the star particles with r-band luminosity weighting. Ellipticities are derived from the eigenvalues of the 2D luminosity weighted inertia tensor of the star particles (positions are projected on a plane orthogonal to the line of sight). All star particles within one half-luminosity radius are used in the calculation. All other quantities are computed on mock spaxels. For each galaxy, its star particles are projected on a plane orthogonal to the line of sight and sorted in a two dimensional spatial grid covering all star particles within one half luminosity radius, with a fixed pixel width of 1.5 kpc. In each pixel, we compute the average r-band luminosity, and we fit the average line-of-sight velocity and dispersion. We then compute the luminosity weighted average over all the pixels with more than 10 particles. The \ha\ sample contains 30,475 galaxies with stellar mass greater than $\mstar=5\times10^9\msun$. 

\subsection{MAGNETICUM Simulations}
\label{subsec:magneticum_sims}

The third set of cosmological hydrodynamical simulations that we will use are the \textit{Magneticum Pathfinder}\footnote{www.magneticum.org} simulations, hereafter simply \ma\ (see Dolag et al., in prep and \citealt{hirschman2014} for more details on the simulation). We use the data from the medium-sized cosmological box (Box 4) with a volume of (59 Mpc)$^3$ co-moving at the ultra high resolution level. \ma\ adopts a cosmology compatible with the Wilkinson Microwave Anisotropy Probe 7 cosmology \citep[$\Omega_\mathrm{m}$=0.272, $\Omega_{\Lambda}=0.728$, $H_{0}=70.4$\kms\ Mpc$^{-1}$;][]{komatsu2011}. The dark matter and gas particles have masses of respectively $3.7\times10^7\msun$ and $7.3\times10^6\msun$, and each gas particle can spawn up to four stellar particles.

Structural, stellar population, and stellar kinematic measurements are described in \citet{schulze2018}, here corrected to $H_{0}=70.0$\kms\ Mpc$^{-1}$. However, here we use luminosity weighted quantities as compared to the mass-weighted quantities in \citet{schulze2018}. $r$-band luminosities are derived using an identical method that was used for the \ea\ data. The effective radius of a galaxy is estimated from the 2D projected $r$-band luminosity map. Kinematic maps are constructed from the mean projected velocity along a line-of-sight and the velocity dispersion is derived from the standard deviation of the particle velocities. Similar to some observational surveys (e.g., \at, CALIFA), Voronoi tessellation is adopted to avoid low-numbers of particles along a line-of-sight. Ellipticities for a given projection are derived following the definition of \citet{cappellari2007}, but using an iterative process. First, $\epsilon$ is calculated from all particles within a circular aperture radius of 1.5\re. Then this process is repeated using a new aperture with the previously determined ellipticity $\epsilon$ containing the same stellar mass, until the estimate for $\epsilon$ converges. This technique differs slightly from the approach taken for \ea\ and \ha, where no iterative approach was used. Stellar ages were calculated as an $r$-band luminosity weighted stellar age using all particles within $r_{50}$. Because the dark matter particle mass in \ma\ is significantly higher than in \ea\, we adopt a higher mass limit of $\mstar=1\times10^{10}\msun$. Above this stellar mass limit, the \ma\ sample contains 2073 galaxies.

\subsection{Summary of Data from Simulations}
\label{subsec:summary_sims}

We have constructed a large sample of galaxy mock-observations from four major hydrodynamical cosmological simulations: \ea\ and \hy, \ha, and \ma. In all simulations, we have used measurement techniques as close to observational measurements techniques as possible, but minor differences still exist. Furthermore, we use simulations with different particle resolution and with different volume sizes: in \ea\ the dark particle mass is $m_{\rm{dm}}=9.7\times10^6\msun$ in a (100 Mpc)$^3$ co-moving volume, for \ha\ we have $m_{\rm{dm}}=8\times10^7\msun$ and (142 Mpc)$^3$ co-moving volume, and for \ma\ $m_{\rm{dm}}=3.7\times10^7\msun$ in (59 Mpc)$^3$ co-moving volume. We stress that these simulations adopt different philosophies for calibrating to and reproducing observational results. Where some are ``made to match", others are ``made to bridge" (i.e., calibrated on larger-scale observed statistical properties, versus calibrated on ensemble average smaller-scale simulations). Therefore, our main aim of the paper is to identify key areas of success and tension, not to determine which simulations provides the closest or "best" match to observations.

\section{Comparing Observations and Simulations}
\label{sec:comp_obs_sim}

In the following section we will compare fundamental galaxy parameters obtained from simulations and observations using similar measuring techniques. We will look at structural parameters, such as effective radius and ellipticity, but also at dynamical properties like aperture velocity dispersion and \vse, the average ratio of the velocity and velocity dispersion within one effective radius. The main goal of this section is to see where simulations and observations agree or disagree. Section \ref{sec:previous} and \ref{sec:summary_discussion} will be devoted to understanding where the differences are coming from and what lessons can be learned from this.


\begin{figure*}
	\includegraphics[width=\linewidth]{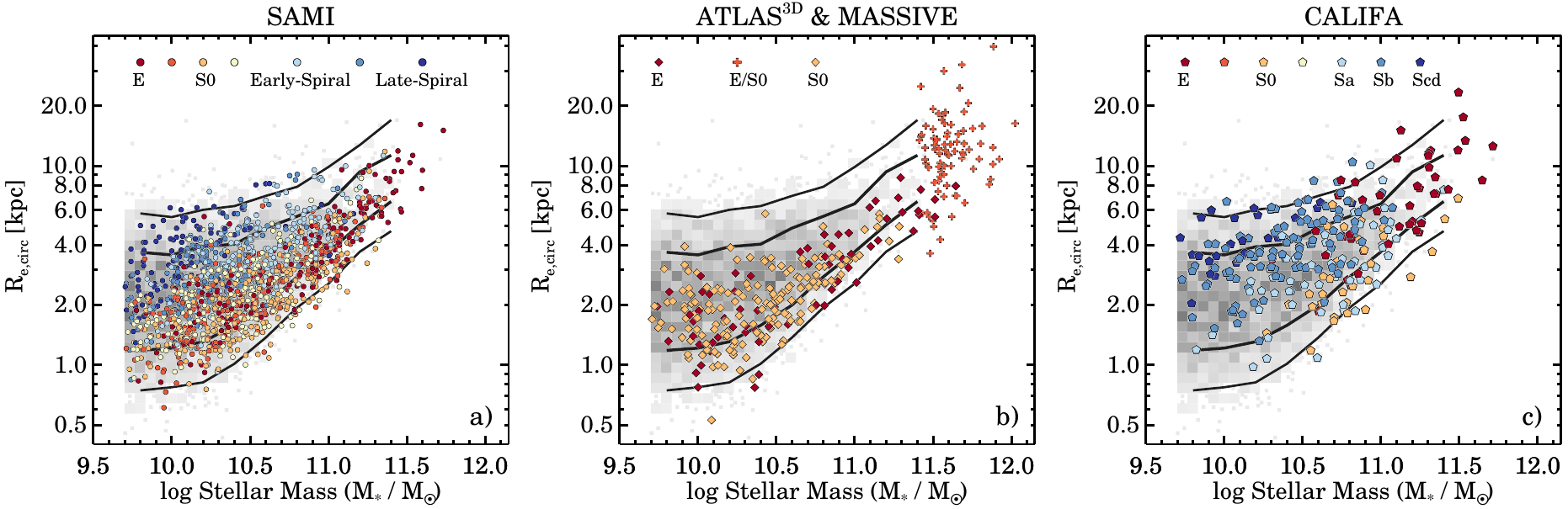}
    \caption{Comparison of GAMA and IFS observational data in the size-mass diagram. We show the density of galaxies from the GAMA survey as grey squares, where darker grey means higher density of galaxies. The grey lines show the 2.5th, 16th, 84th, and 97.5th percentile of the GAMA data in bins of fixed stellar mass with a width of 0.2 dex. The individual symbols show the IFS data from the SAMI Galaxy Survey (left), \at(middle; diamond symbols) and MASSIVE (middle; plus symbols), and CALIFA (right). IFS data are coloured coded by their visual morphological classification as indicated by the legends. We recover the well-known morphology trend where below $\logm\sim11$ early-type galaxies tend to be smaller than late-type galaxies, whereas early-type galaxies start to match the sizes of late-type galaxies at the highest stellar masses ($\logm>11.3$).}
    \label{fig:mass_size_samples}
\end{figure*}


\subsection{Observational Biases}
\label{subsec:obs_bias}

Most observational selection effects are well understood and easily reproducible, but the combination of four different surveys with four different sets of selection criteria makes the comparison with the simulations challenging. There is not a single, simple selection function that encompasses all the biases of the observations. The strongest observational bias is as a function of stellar mass: as galaxies decrease in stellar mass their total luminosity decreases, which makes it increasingly hard to obtain the targeted S/N per galaxy spaxel. Therefore, for the sake of simplicity, we only apply a stellar-mass selection to the simulated data. This will remove the strongest bias, but more subtle effects may arise from the other selection criteria (for details see Section \ref{sec:obs_data}). Note, for example that \cite{canas2018} find an offset between the stellar mass function from observations \citep{moustakas2013, wright2017} and Horizon-AGN \citep[see also][]{kaviraj2017}. This implies that with our mass-matching technique, on average low-mass galaxies in \ha\ may be selected in lower-mass halos than in \ea. For \ma\, a comparison of the stellar-to-halo mass relation was made in \citet{teklu2017}, and their simulated galaxies agree qualitatively with the different observations. However, possible biases might arise in \ma\ due to the smaller volume box used that could lead to lower density environments being probed on average compared to \ea, \ha\ and the observations. As the evolution of the angular momentum in galaxies is sensitive to halo growth, this could add to the biases in the comparison between the simulated and observed dynamical measurements.

To reveal possible observational biases as a function of stellar mass, in Fig.~\ref{fig:mass_size_samples} we present the size-mass relation of the observational samples compared to an unbiased sample from GAMA at $z<0.1$. The data are colour coded by visual morphology. Note that the MASSIVE sample only contains ellipticals (E) and lenticulars (S0), however, no individual galaxy classifications were available to us.

\begin{figure*}
\includegraphics[width=\linewidth]{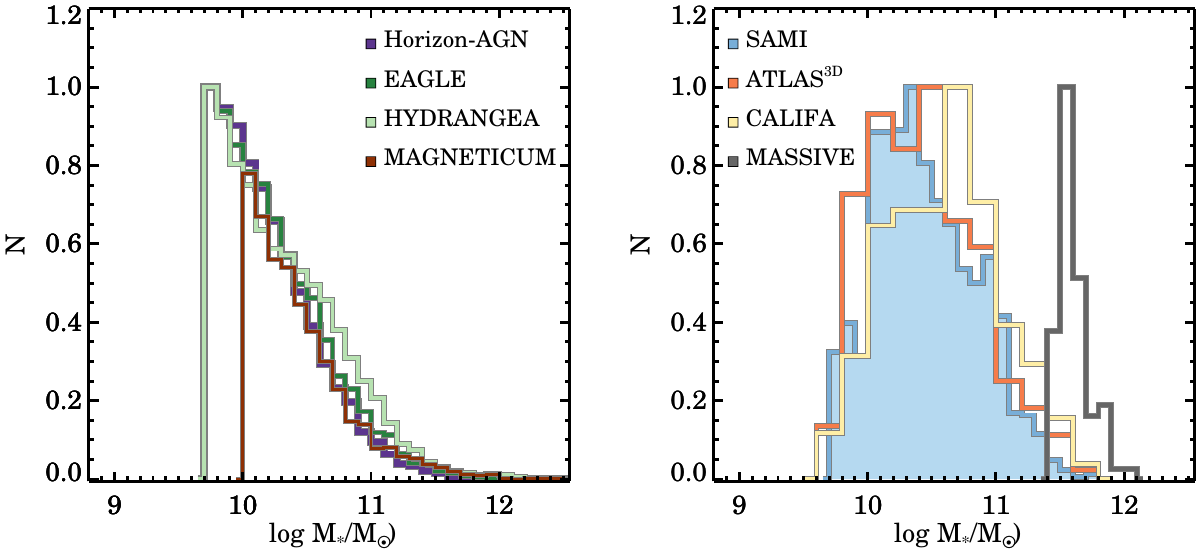}
\caption{Normalised distribution of galaxy stellar masses from simulations (left panel) and observations (right panel). Simulation data from \ha\ are shown in purple, \ea\ in green, \hy\ light-green, and \ma\ in brown. Observational data from SAMI is shown in blue, \at\ in orange, CALIFA in yellow, and MASSIVE in grey. Because the SAMI sample is the largest observational sample, for clarity the histogram is shown with a colour fill. Due to observational limitations and selection criteria, the number of galaxies below $\logm<10$ rapidly decreases in the observational surveys. To overcome this bias as a function of stellar mass, data from simulations are mass-matched to the observational data (see Section \ref{subsec:mass_dist}).}
\label{fig:mass_dist}
\end{figure*}

A size-mass dependence on morphology is well known; at lower stellar mass ($\logm<10$) early-type galaxies tend to be smaller than late-type galaxies \citep[e.g.,][]{shen2003,lange2016}. The early-type size-mass relation is steeper as compared to the relation for late-type galaxies, and at high stellar masses ($\logm>11$) early-type galaxies start to become larger. This trend with morphology is clearly visible in the SAMI Galaxy Survey data, in particular between $10<\logm<11$. Similarly, because the \at\ survey consists of early-types only, galaxies are on average smaller than the GAMA galaxies.

Below $\logm<10.5$, the fraction of late-type galaxies strongly increases. Below $\logm<10$ we find a bias towards more compact galaxies, as there are few SAMI galaxies above the 1-$\sigma$ GAMA contour line around 4kpc. The CALIFA sample becomes strongly dominated by late-type galaxies at $\logm<10.5$ and CALIFA galaxies are on average slightly larger than the GAMA comparison sample at this stellar mass. The sizes of the MASSIVE galaxies appear to lie well above the different surveys. We will discus this further in Section \ref{subsec:mass_mdyn}.

While individual surveys show small morphological biases, we can diminish the effect of these biases by combining the data from the SAMI Galaxy Survey, \at, and CALIFA into one sample. We note that combining the different surveys does not completely remove the sample biases, but lessens the biases of the individual surveys.

\subsection{Mass Distribution and Sample Matching}
\label{subsec:mass_dist}

We now describe the mass selection of simulated galaxies in order to match the observational sample. We start by showing the normalised mass distributions of the simulated and observational data in Fig.~\ref{fig:mass_dist}. Note that due to the different stellar mass range of the \ma\ sample, we have normalised the stellar mass distribution of \ma\ to have a peak of 0.78, rather than 1.0 that was adopted for \ea, \hy, and \ha. This way, we can better compare the shape of the distribution between the different simulations. 

We find a close match between the mass distribution of \ea, \ha, and \ma\ (left panel), although the \ha\ distribution is slightly higher at low stellar mass ($\logm<10.5$). The clear difference between the \ea\ and \hy\ stellar mass distribution is due to the \hy\ cluster environment. For the observations (right panel), we see that the shape of the distributions for SAMI and \at\ closely match, with a median stellar mass of \logm=10.4. The CALIFA distribution is skewed towards slightly higher stellar mass (median \logm=10.6), and the MASSIVE survey is a clear but intended outlier, with a narrow distribution at very high-stellar mass (median \logm=11.6).

As the simulated and observed stellar mass functions differ in shape, we cannot directly compare parameters from the simulations and observations without introducing a bias caused by trends with stellar mass. Therefore, to remove this bias, we will perform a mass-matching by randomly selecting simulated galaxies as a function of stellar mass. The number of galaxies in the \ea, \hy, and \ha\ simulations exceeds the number of galaxies in the observed sample. Thus, we can select many more galaxies from these simulations than there are in the observed sample. However, in \ma, the number of galaxies is less than in the observed sample, and in order to mass-match the sample, we have to select an even lower number of galaxies.

Despite the differences in model parameters between \ea\ and \hy, we combine the two models because it was shown in \citet{lagos2018b} that both models have a similar span in the \lr\ (spin-parameter proxy) - ellipticity space, and the differences were due to stellar mass sampling. In observations, the ratio of field and group galaxies compared to cluster galaxies in the observations is approximately 40 percent. As mentioned in Section \ref{subsec:eagle_sims}, in order to match this number in the \ea\ and \hy\ simulations, we only select \hy\ galaxies that are in groups or clusters with mass greater than $\logm_{\rm{group}}>13.85$. From now on, we will refer to this joined \ea\ and \hy\ sample as \eap.

For the mass-matching, we first determine the number of galaxies in both the simulated and observational datasets, in stellar mass bins with a 0.15 dex width, starting at $\logm=9.7$. In every mass-bin we then compare the number of simulated versus observed galaxies, and calculate the ratio between the two. For example, in the stellar mass bin of $10.0<\logm<10.15$, we find 1227 \eap\ galaxies, whereas the combined observational dataset contains 277 galaxies. Thus the ratio of simulated to observed galaxies is a factor of 4.43. Over the entire stellar mass range, we find that the lowest ratio of simulated to observed galaxies is 2.00 in the stellar mass range of $10.9<\logm<11.05$ (320 and 160 galaxies in \eap\ and observations, respectively). Similarly, for \ha\ the lowest ratio is 4.25 at $11.2<\logm<11.35$, and for \ma\ the lowest ratio is 0.52 at $10.9<\logm<11.05$

The lowest ratio then sets the number of galaxies that we can randomly sample in each mass bin from the simulations. That is, for \emph{every} mass bin in the \eap\ (\ha, \ma) simulation, we randomly select 2.00 (4.25, 0.52) times as many galaxies as there are observed. This way, the mass-distribution of the simulations will be identical to the observations, while also maintaining the largest number of simulated galaxies to which to compare to. As we only do a single random draw, the choice of the random seed may potentially impact our results when the sample to draw from gets small (e.g., towards high stellar mass). We check and confirm that by using different random seeds none of our our conclusions change.

We set an upper limit for the mass-matching at $\logm=11.5$, where the number of simulated galaxies is low as compared to the observations. Above a stellar mass of $\logm=11.5$, there are 82 galaxies in the combined surveys including MASSIVE, while there are 109 in \eap, 166 in \ha, and 38 in \ma. If we were to include the MASSIVE sample in the mass matching, the overall sampling factor would be $\sim2$ or less. Thus, to keep the sampling factor as high as possible for the combined observed sample we limit the mass matching to $\logm<11.5$. Effectively, this means that we combine the SAMI, \at, CALIFA, and one fifth of the MASSIVE measurements into one sample. 

Combining the different surveys also helps to homogenise coverage for several observed parameters (see e.g., Fig.~\ref{fig:mass_size_samples}). We note that our observational sample is biased towards early-type galaxies. Mass-matching will not correct for possible morphological differences between the observed and simulated samples, but is beyond the scope of this paper to try and correct for such second order effects. Finally, because the MASSIVE Survey sample only contains galaxies with stellar mass $\logm>11.5$, we will describe the comparison with the MASSIVE sample separately in the following sections.

\begin{figure*}
\includegraphics[width=\linewidth]{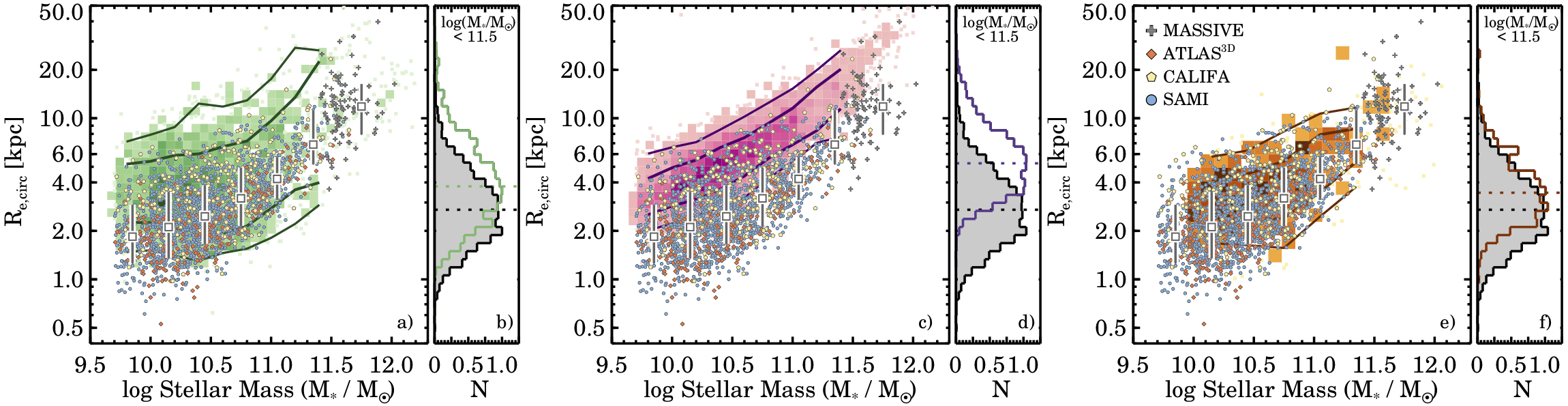}
\caption{Comparison of simulated and observational data in the size-mass diagram. In panel a), we show the density of galaxies from the combined \ea\ and \hy\ simulations as green squares, where darker green means higher density of galaxies. The green lines show the 2.5th, 16th, 84th, and 97.5th percentile of the data in bins of fixed stellar mass with a width of 0.2 dex (0.3 dex for \ma\ data). Observational data from the SAMI Galaxy Survey is shown as blue circles, \at\ as orange diamonds, CALIFA as yellow pentagons, and MASSIVE as grey pluses. The grey and white squares show the median of the observed sample in mass bins, and the vertical lines show the 16th and 84th percentile of the observed distribution. In panel b) we show the effective radius-distribution of \eap\ (green) and observational data (black), where the dotted lines show the median in effective-radius. Data from the MASSIVE survey is not included in this panel (see Section \ref{fig:mass_dist}). The \eap\ size-mass relation matches reasonably well with observational data, although the median size of simulated galaxies is larger by 42 percent as compared to our observed sample (median of 3.82 kpc versus 2.67 kpc, respectively). For \ha\ (panel c-d), the size-mass relation is offset from observational data; at all stellar masses simulated galaxies are too large. The median size is 5.26 kpc as compared to 2.67 kpc for the observations, but the offset is similar at all stellar masses. The \ma\ size-mass relation (panel e-f) has a similar slope and spread, but the average galaxy size is slightly too large (median of 4.68 kpc versus 3.45 kpc in observations), except for the most massive galaxies.
}
\label{fig:mass_size}
\end{figure*}

\subsection{Size-Mass}
\label{subsec:mass_size}

We start the comparison between observations and simulations with one of the most fundamental galaxy relations, between the stellar mass and effective radius (Fig~\ref{fig:mass_size}). Due to the availability of high-spatial resolution \textit{Hubble Space Telescope} imaging campaigns \citep[e.g., COSMOS, CANDELS, etc;][]{capak2007,koekemoer2007,grogin2011,koekemoer2011}, the size-mass relation is extremely well studied out to high-redshift and offers invaluable insight into how galaxies grow over time \citep[e.g.,][and references therein]{vanderwel2014}.

The match between \eap\ and observations is good (Fig.~\ref{fig:mass_size}a), but \eap\ galaxies have sizes that are on average too large. While the \ea\ simulations are not directly calibrated to match disk-type galaxies, models were rejected that produced galaxies that were far too small \citep{crain2015}, so a good match here is not entirely unexpected \citep[see][]{schaye2015}. However, the calibration was performed with mass-weighted sizes, which are typically smaller than luminosity-weighted (see Appendix \ref{sec:app_lum_mass}). The addition of the \hy\ data increases the number of early-type galaxies, and thus adds more data to the bottom-right side of the size-mass relation. Nonetheless, at fixed stellar mass, we observe the spread in the \eap\ sizes to be larger as compared to the observed sample. In Fig.~\ref{fig:mass_size}b, we find an offset between the median effective radius of the simulations versus the observations (3.82 kpc versus 2.67 kpc, respectively). From bootstrapping the distributions a thousand times, we find that the uncertainty on the medians are small (1-2 percent). The number of observed galaxies is largest between $10<\logm<10.5$, which is also where the observational bias towards early-type galaxies is the largest. Thus part of the mismatch here may be ascribed to a morphological observational bias. At the highest stellar masses ($\logm>11.5)$, the match to the MASSIVE data is excellent.

In Fig.~\ref{fig:mass_size}c-d we present the size-mass relation from the \ha\ simulations. We find a good qualitative match between the shape and slope of the size-mass relation, but the simulated size-mass relation is offset towards larger radii as compared to observations by a factor of 1.88 (median effective radius of the simulations versus the observations is 5.26 kpc versus 2.67 kpc, respectively). The spread in size at fixed stellar masses is significantly smaller as compared to \ea, but also slightly smaller as compared to observations. The offset towards larger radii is not likely caused by a morphological bias in the simulations towards disk-type galaxies. The offset is similar at all stellar masses, and in \ha\ the size-mass relation for early-type lies above the relation for late-type galaxies at all stellar masses \citep{dubois2016}. The shape of the size-mass relation is surprisingly similar for early-type and late-type galaxies in \ha. We will discus this offset further in Section \ref{sec:previous}.

We investigate the size-mass relation from the \ma\ simulations in Fig.~\ref{fig:mass_size}e-f. Similar to both \eap\ and \ha\, the shape and slope of the size-mass relation are in good agreement with the observations, but for \ma\ there is only a minor offset between the observed and simulated average size, with the \ma\ sizes being sightly larger. The spread of the distribution is similar to the observed sample. We find an excellent agreement between the \ma\ and the MASSIVE sample (i.e., at $\logm>11.5)$. In Fig.~\ref{fig:mass_size}f, the median offset between the effective radius of the simulations versus the observations is 4.68 kpc versus 3.45 kpc, respectively. 

In conclusion, the size-mass relation is qualitatively well-reproduced by all simulations, but on average simulated galaxies are too large. We checked whether our conclusion would change if we use the full GAMA dataset (see Figure \ref{fig:mass_size_samples}), rather than the combined IFS sample, but the results remain the same. The spread in sizes at fixed stellar masses varies between different simulation, from larger to slightly smaller as compared to observations.


\begin{figure*}
\includegraphics[width=\linewidth]{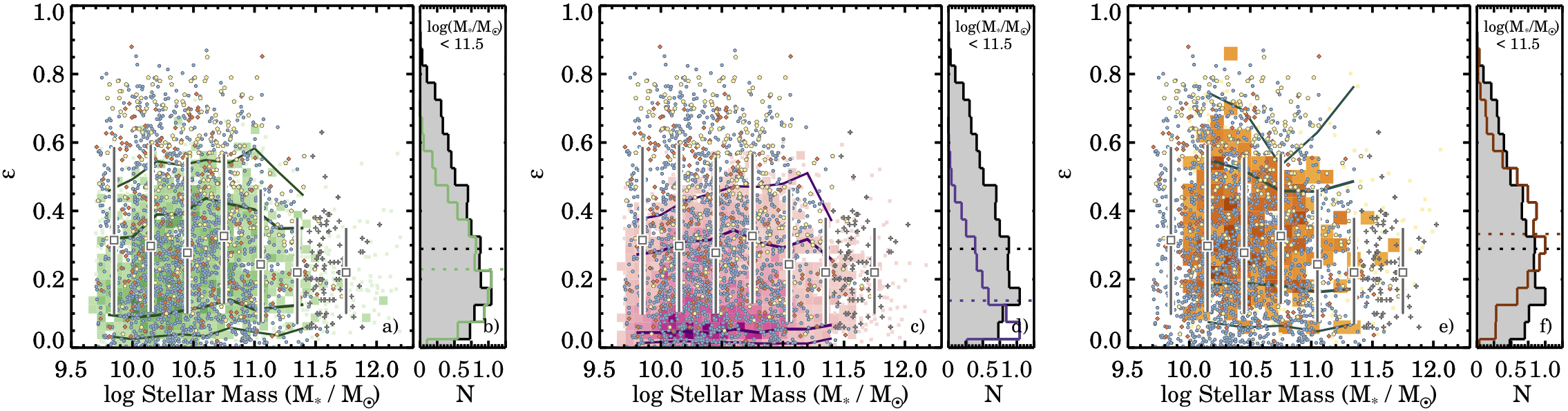}
\caption{Comparison of simulated and observational data in the ellipticity-mass diagram. Symbols are the same as in Fig.~\ref{fig:mass_size}. We find a good match between the ellipticities from \eap\ and observations, although there are too few flattened high $\epsilon$ galaxies in \eap. Simulated \ha\ galaxies are too round by a large factor, except for the highest stellar masses. Almost no \eap\ galaxies with $\epsilon>0.6$ or \ha\ galaxies with $\epsilon>0.5$ exist. At low stellar masses, \ma\ provides a significantly better match at high $\epsilon$ as compared to the other simulations, yet the number of observed round galaxies $\epsilon>0.2$ is lower as compared to the observed sample.
}
\label{fig:mass_ellipticity}
\end{figure*}


\subsection{Ellipticity-Mass}
\label{subsec:mass_ellip}

In Fig.~\ref{fig:mass_ellipticity} we compare the relation between the observed ellipticity and stellar mass. The overall shape of the distribution, where the most massive galaxies become increasingly round, is well-recovered by \eap, \ha, and \ma. For both observations and simulations we find that above $\logm\sim11$ flat galaxies start to disappear. Above this stellar mass limit, galaxies formed by dry-major mergers start to dominate the galaxy population \citep[e.g.,][]{cappellari2016}, and hence we observe fewer flat galaxies. However, above $\logm>11.5$, galaxies in both \eap\ and \ha\ appear to be rounder (median $\epsilon=0.18, 0.17$) as compared to the MASSIVE galaxies (median $\epsilon=0.23$), whereas \ma\ produces a significant number of extremely massive galaxies with $\epsilon>0.25$). However, as the ellipticities for MASSIVE galaxies are not derived within one effective radius but from a global isophote, this offset between \eap\, \ha\, and MASSIVE, could possibly be attributed to a difference in measuring technique. 

On average, the \eap\ ellipticities are slightly lower (rounder) than our observed sample. The offset is most pronounced below $\logm<11$, where there are almost no galaxies with $\epsilon>0.6$. The median ellipticity is $\epsilon=0.23$ for \eap\ versus $\epsilon=0.29$ in the observed sample, but the 97.5 percentile $\epsilon$ values of the distribution are 0.54 and 0.74, respectively. For \ha\ the offset is more dramatic, the median ellipticity $\epsilon=0.14$ which is a factor 2.0 lower than observed, and the 97.5th percentile $\epsilon$ value is 0.45. \ma\ is the only simulation that produces flattened galaxies with ellipticities as high as detected in the observations (median $\epsilon=0.33$, 97.5 percentile $\epsilon$=0.69). However, the \ma\ has a notable lower number of extremely round galaxies at all stellar masses.

\citet{lagos2018b} showed that the missing high-ellipticity galaxies in \ea\ was not due to the resolution of the simulation, as the analysis of the higher resolution \ea\ runs did not improve the sampling of the high ellipticity range. The authors therefore concluded that this is a limitation of the ISM model and cooling adopted in \ea. The ISM model includes a temperature floor of 8,000K, i.e., gas is not allowed to cool below this temperature. This temperature limit corresponds to a Jeans length of approximately 1kpc physical, and thus, disks cannot be thinner than this. Our Milky Way and other local disks, however, display thinner disks with typical scale heights of 300-700pc \citep[e.g.,][]{kregel2002,blandhawthorn2016}. Thus, an important improvement needed in simulations to reproduce the very high ellipticity galaxies, is to realistically model the ISM to produce more realistic vertical structure of disks. Because \ha\ employs a similar ISM model, we expect the reasoning above to also apply to this simulation. Nonetheless, it should be noted that \ha\ has a slightly lower spatial resolution than \eap\ and that, due to the method used to compute the hydrodynamics, this spatial limit represents a sharp scale cut where it is no longer possible to derive gradients in the gas properties, at fixed positions. This is also the minimal scale on which feedback energy and momentum from supernovae and AGN feedback can be released (usually several times that to distinguish anisotropic and isotropic modes of AGN feedback for instance). Such processes therefore limit the formation of thin, highly rotating discs, including at high stellar mass in \ha.

\begin{figure*}
\includegraphics[width=\linewidth]{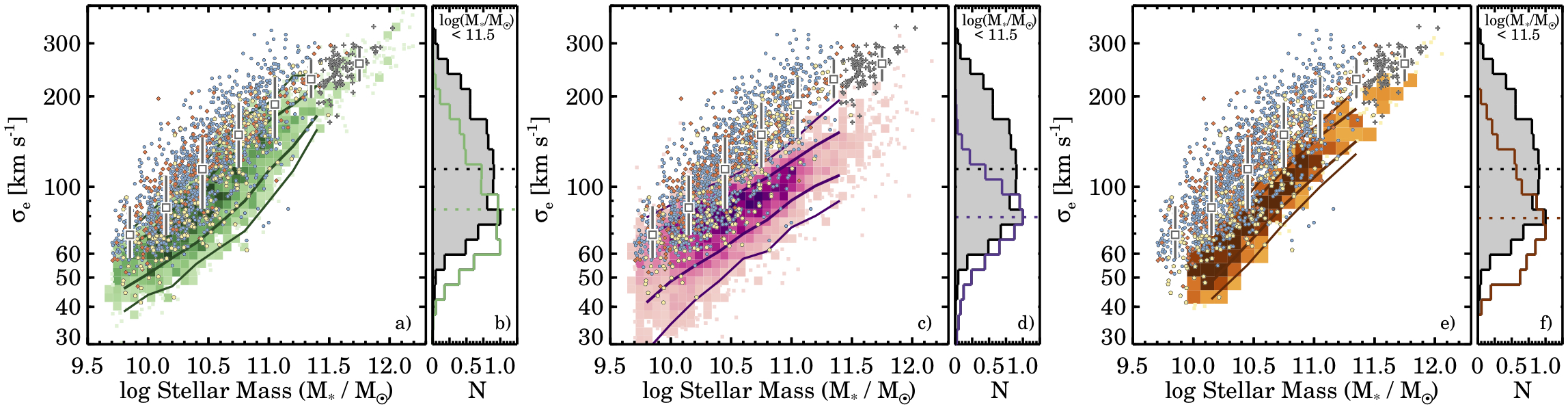}
\caption{Comparison of simulated and observational data in the stellar velocity dispersion-mass diagram. Symbols are the same as in Figure \ref{fig:mass_size}. While the shape of the \eap\ $\sigma$-mass is the same as observation (panel a), the median velocity dispersion is lower: 83.5 \kms\ versus 114.6 \kms\ in the observational data (panel b). The \ha\ relation has a different coefficient and does not extend to high-velocity dispersions (panel c). The median velocity dispersion in \ha\ is lower than in the observations (panel d, 78.9 \kms\ versus 114.6 \kms\, respectively). The shape of the \ma\ $\sigma$-mass is the same as observations (panel e), but the relation is offset to lower velocity dispersion (78.6 \kms\ versus 114.6 \kms\ in the observational data panel f).
}
\label{fig:mass_sigma}
\end{figure*}

\subsection{Velocity Dispersion-Mass}
\label{subsec:mass_sigma}

We will now include dynamical information in the comparison. The first step is to look at aperture velocity dispersions that are also accessible using single fibre spectroscopy. However, by using IFS observations we have the advantage that we can precisely define our aperture to match the effective isophote of the galaxies and thus remove any aperture-related uncertainties that might arise from a fixed fibre size. We use a flux-weighted sum of all velocity dispersion measurements within one effective radius:

\begin{flushleft}
\begin{equation}
\sigma_{\rm{e}}^2 = \frac{ \sum_{i=0}^{N_{spx}} F_{i}\sigma_{i}^2}{ \sum_{i=0}^{N_{spx}} F_{i}}.
\label{eq:ss}
\end{equation}
\end{flushleft}

\noindent Here, the subscript $i$ refers to the position of a spaxel within the ellipse with semi-major axis \re\ and ellipticity \ee, $F_{i}$ the flux of the $i^{th}$ spaxel, $\sigma_{i}$ the velocity dispersion in \kms. Not all SAMI and \at\ measurements have coverage out to one effective radius, and we only include galaxies with a minimum fill factor of 85 per cent. 

Note that this approach differs from a single-fibre or composite spectrum velocity dispersion measurement. To match a single-fibre measurement, we first need to sum all spectra within an aperture and then measure the stellar velocity dispersion from that spectrum. While it has been shown that \sqvrmse\ closely approximates the single-fibre velocity dispersion measurement (see Section \ref{subsec:mass_mdyn} Eq. \ref{eq:vrms}), here our aim is to focus on the velocity dispersion without including the rotational velocity $V$. For the MASSIVE survey data the velocity dispersion maps are not available. Instead, we use the tabulated data from \citet{veale2017a}.

We present the velocity dispersion (\se) - stellar mass relation in Fig.~\ref{fig:mass_sigma}. The observational data show a clear increasing trend between $\sigma$ and stellar mass, but there is a noticeable turnover around $\logm\sim11.3$. This turnover will be further discussed in the next Section \ref{subsec:mass_mdyn}, where we compare dynamical and stellar mass estimates.

The shape of the \eap\ data matches the observed relation well, but the simulated data are lower by 0.15 dex at fixed stellar mass. At the highest stellar masses, there is an excellent agreement between the MASSIVE and \eap\ velocity dispersions. From Fig.~\ref{fig:mass_sigma}b we find that the width of the distribution of \se\ values from the \eap\ simulations is similar, but slightly more skewed towards lower \se\ values as compared to observations (0.18 dex versus 0.16 dex in simulations and observations respectively). At fixed stellar mass, between $10.4<\logm<10.6$ the scatter in \eap\ is slightly smaller than observations (0.09 dex versus 0.11 dex, respectively).

One might naively expect that this offset could be caused by observational biases due to seeing that are known to increase the stellar velocity dispersion as the absorption lines get artificially broadened by the velocity profile of the galaxy. However, as the SAMI and \at\ data are similar, this indicates that seeing is not a likely cause for the higher velocity dispersions in the data as compared to observations.

The closest match between \ha\ and observations is at low stellar masses, but overall the slope of the \ha\ relation is shallower and the velocity dispersions are too low (Fig.~\ref{fig:mass_sigma}c). Below $\logm=10.5$ the offset is approximately 0.12 dex whereas, at high stellar masses, the velocity dispersions are too low by 0.23 dex. From Fig.~\ref{fig:mass_sigma}d we also find that the width of the distribution is smaller in \ha\ as compared to observations (0.08 dex at $10.4<\logm<10.6$). However, this is due to a different slope in the relation from the simulations, and the fact that \ha\ has few high-dispersion galaxies.

In Fig.~\ref{fig:mass_sigma}e we show the results from the \ma\ simulations. The shape of the \ma\ \se-mass relation matches the observations well. However, the spread in \se\ at fixed stellar mass ($10.4<\logm<10.6$) is significantly smaller as compared to observations (0.05 dex versus 0.11 dex, respectively). Similar to \eap\ and \ha, the \ma\ velocity dispersions are also offset to lower values as compared to the observed \se-mass relation. We note that none of the simulations show the strong change in the slope of the \se-mass relation around a stellar mass of $\logm\sim11.3$.

\begin{figure*}
\includegraphics[width=\linewidth]{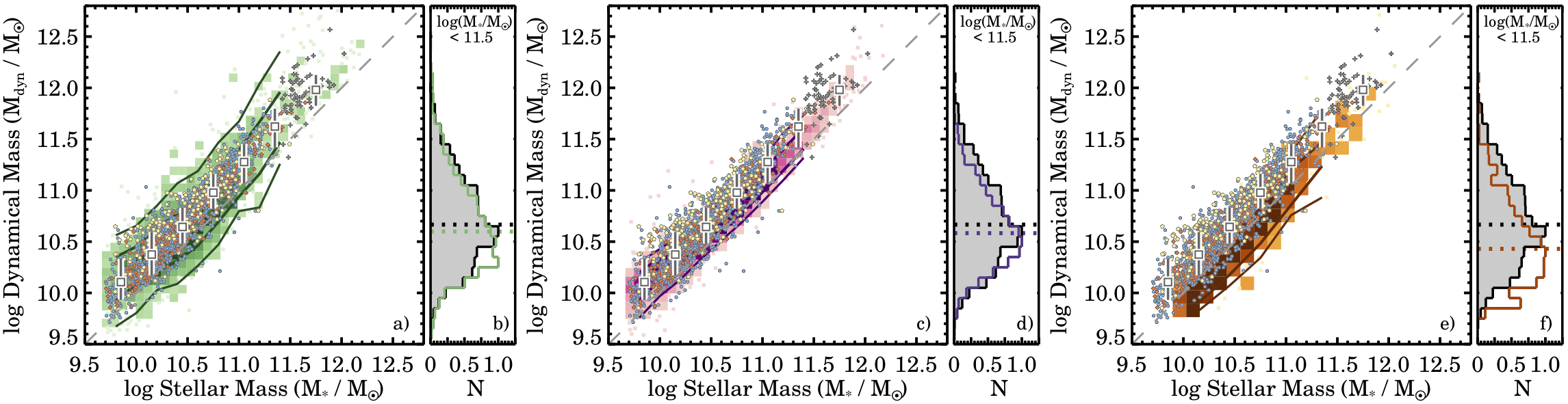}
\caption{Comparison of simulated and observational data in the dynamical-stellar mass diagram. Symbols are the same as in Figure \ref{fig:mass_size}. The combination of size and velocity dispersion largely reconciles the discrepancy between observations and simulations that were seen in Fig.~\ref{fig:mass_size} and Fig.~\ref{fig:mass_sigma}.}
\label{fig:mass_mdyn}
\end{figure*}


\subsection{Dynamical-Stellar Mass}
\label{subsec:mass_mdyn}

We now calculate dynamical masses from the size and the projected second velocity moment using the Virial theorem, to reconcile the offsets that we found between the observational and simulated size-mass and \se-mass relations. Secondly, in the observational data we also noted that at the highest stellar masses, the size-mass relation becomes steeper, whereas the velocity dispersion-mass relation becomes less steep. As this turnover coincides with the addition of the MASSIVE data, one might worry that there are remaining observational biases between the different samples. To address these issues, we will now compare dynamical and stellar mass estimates.

Dynamical masses are derived from the projected second velocity moment \vrmse\ and the circularised effective radius using the following expression:
\begin{equation} 
M_{\rm dyn}=\frac{\beta~ \vrmse~R_{\rm e, c} }{G}.
\label{eq:mdyn}
\end{equation}
Here, \vrmse\ is defined as:
\begin{flushleft}
\begin{equation}
\vrmse = \langle V^2 + \sigma^2 \rangle_{\rm{e}} \equiv \frac{ \sum_{i=0}^{N_{spx}} F_{i}(V_{i}^2+\sigma_{i}^2)}{ \sum_{i=0}^{N_{spx}} F_{i}}.
\label{eq:vrms}
\end{equation}
\end{flushleft}
For the dynamical mass calculation we use \vrmse\ rather than $\se^2$ from Eq.~\ref{eq:ss} as it was shown by \citet{cappellari2013a} and \citet{veale2017a} that \vrmse\ is a better approximation of the composite derived $\se^2$ that is used for deriving dynamical masses \citep[see also][]{cappellari2006}. For the MASSIVE survey no \vrms\ measurements are available. However, \citet{veale2017a} showed that their \se\ values closely approximate both the composite \se\ and \sqvrmse\ for all slow rotators in their sample. Hence, for the MASSIVE sample, we use the tabulated \se\ data from \citet{veale2017a}, but we exclude all fast rotatoring galaxies in this section.

We use a $\beta(n) = 5$ that is commonly used for early-type only samples \citep{cappellari2006}. $\beta$ can also be derived from an analytic expression as a function of the S\'ersic index, as described by \citet{cappellari2006}, but this is beyond the scope of the paper as S\'ersic index measurements are not available for the full sample.

In Fig.~\ref{fig:mass_mdyn} we compare the observational and simulated data in the $M_{\rm dyn}-\mstar$ plane. First, we note that the large majority of observational data lie above the one-to-one relation, but that their offset from this relation increases with increasing stellar mass. This increase can be explained by increasing non-homology, dark matter fraction, and possibly a varying IMF \citep[e.g.,][]{vandokkum2010,cappellari2012b}. While the exact nature of this offset is beyond the scope of this paper, we do note that there is no sudden change in the $M_{\rm dyn}-\mstar$ relation around a stellar mass of $\logm\sim11.3$. By comparing dynamical and stellar masses we can now conclude that there are no measurement biases in \re\ and \se\ from the difference observational samples. Instead, above a stellar mass of $\logm\sim11.3$, massive-round slow rotating galaxies formed by dry-major mergers start to dominate the galaxy population (for a recent review on this topic see \citealt{cappellari2016}, but also \citealt{robotham2014}).

The combination of size and velocity dispersion also largely reconciles the discrepancy between observations and simulations. This is important to note, as we will continue to compare both the rotational and random motions of stars in the next section. \eap\ and \ha\ are in good agreement with the observations, both in terms of the median and the spread of the $M_{\rm dyn}-\mstar$ distribution (Fig.~\ref{fig:mass_mdyn}a,c), whereas \ma\ is offset to the lower-right. The distribution of dynamical masses in Fig.~\ref{fig:mass_mdyn}b,d is lower by approximately $-0.08$ dex in the simulations as compared to observations, but larger ($-0.24$ dex) for \ma\ in \ref{fig:mass_mdyn}f. While the exact cause for this offset is unclear, as mentioned above, there are many factors that could contribute to this relatively minor offset.

\subsection{\vs-Mass}
\label{subsec:mass_vsigma}

In the following section we will look at the \vse\ measurements. \vs\ describes the average ratio of the velocity and velocity dispersion within one effective radius. We adopt the following definition by \citet{cappellari2007}:
\begin{flushleft}
\begin{equation}
\left(\frac{V}{\sigma}\right)^2 \equiv \frac{\langle V^2 \rangle}{\langle \sigma^2 \rangle} = \frac{ \sum_{i=0}^{N_{spx}} F_{i}V_{i}^2}{ \sum_{i=0}^{N_{spx}} F_{i}\sigma_i^2}.
\label{eq:vs}
\end{equation}
\end{flushleft}
\noindent Here, the subscript $i$ refers to the position of a spaxel within the ellipse, $F_{i}$ the flux of the $i^{th}$ spaxel, $V_{i}$ is the stellar velocity in \kms, $\sigma_{i}$ the velocity dispersion in \kms. 
For the SAMI Galaxy Survey data, we use the unbinned flux, velocity, and velocity dispersion maps as described in Section \ref{subsubsec:stelkin_sami}. We take the sum over all spaxels $N_{spx}$ that meet the quality cut Q$_1$ and Q$_2$ (Section \ref{subsubsec:stelkin_sami}) within an ellipse with semi-major axis \re\ and axis ratio $b/a$. For the \at\ data, the unbinned flux maps are combined with the Voronoi binned stellar kinematic data as described in Section \ref{subsec:at_data}. The $V$ and $\sigma$ bins are replaced by spaxels with the $V$ and $\sigma$ values belonging to that bin. The approach for CALIFA data is similar to \at\, with the exception of the flux maps that are also Voronoi binned. The kinematic maps for the MASSIVE survey are not publicly available. Instead we convert the spin-proxy measurements \lre\ to \vs\ values using Eq.~B1 from \citet{emsellem2007} assuming $\kappa=1.06$. Note that these measurement are obtained within circularised apertures, but as most MASSIVE galaxies are round with little rotation, the impact of this choice is relatively small. For the simulated \eap, \ha, and \ma\ data, the approach for measuring \vse\ is identical to that of the SAMI Galaxy Survey data.

Not all SAMI and \at\ measurements have coverage out to one effective radius. For all galaxies where the fill factor of good spaxels within one effective radius is less than 95 percent, we apply an aperture correction as outlined in \citet{vandesande2017b}. Note that there is a small number of galaxies with non-regular rotation profiles, for example due to kinematically decoupled cores \citep[e.g.,][]{krajnovic2011}, where the aperture corrections are less applicable. As these galaxies tend to have low \vse\ values, the aperture corrections on \vse\ are typically less than 0.2, and thus do not have a significant impact on our results. A total of 266 SAMI galaxies are aperture corrected (17 percent), 136 for \at\ (56 percent), 21 for CALIFA (8 percent), but none for MASSIVE.


\begin{figure*}
\includegraphics[width=\linewidth]{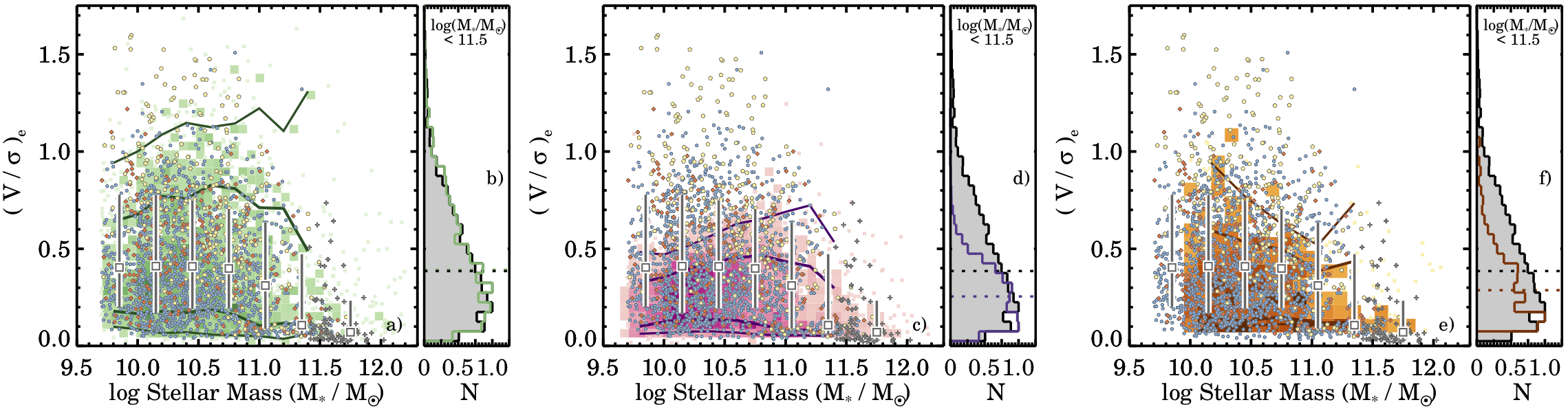}
\caption{Comparison of simulated and observational data in the stellar mass-\vs\ diagram. Symbol are the same as in Figure \ref{fig:mass_size}. There is a remarkably good match between \eap\ and the simulations for \vse\, both as a function of stellar mass and overall distribution. While the trend with stellar mass is correct for \ha, the \vse\ values too low at all stellar mass except $\logm>11.5$. \ma\ galaxies shows a stronger decline with increasing stellar mass as compared to observations, and the overall \vse\ values are lower except at the highest stellar masses.}
\label{fig:mass_vsigma}
\end{figure*}


We present all \vse\ measurements in Fig.~\ref{fig:mass_vsigma}. First, we comment on the variation in \vse\ between the different observational surveys. While the \at\ and SAMI distribution show an excellent agreement, with slightly more high \vse\ late-type galaxies in SAMI, the clear outlier towards high \vse\ is CALIFA. As mentioned in Section \ref{subsec:califa_data}, the isophotal diameter selection of CALIFA creates a bias towards edge-on galaxies, an effect which is also clearly visible in \vse. Due to it's high stellar mass, the MASSIVE sample contains the largest amount of slow-rotating galaxies with low \vse.

There is an excellent match between the observed \vse\ measurements and the \eap\ simulations (Fig.~\ref{fig:mass_vsigma}a-b), whereas \ha\ galaxies on average have too low \vse\ values (Fig.~\ref{fig:mass_vsigma}c-d). In particular there are almost no galaxies that have $\vse>0.7$. Note that pure disks typically have $\vse>0.8$, and thus appear not be produced in the \ha\ simulations. The median \vse\ values of \ma\ galaxies are lower as compared to observations, but in between the range covered by \eap\ and \ha.

All simulations reproduce the observed trend with stellar mass, albeit with small differences. \eap\ and \ha\ show a moderate increase in \vse\ from $\logm=9.5$ to $\logm=10.75$, with a strong decline in \vse\ at $\logm>10.75$, whereas \ma\ galaxies have lower \vse\ with increasing stellar mass across the entire mass range. At the highest stellar masses ($\logm>11.5$), we find that the majority of MASSIVE galaxies are slow rotators with $\vse<0.2$ (median \vse=0.07). In comparison, \eap\ galaxies at the same stellar mass have a considerably higher spin with a median \vse=0.20, and there are several galaxies with $\vse>0.5$. This was also noticed by \citet{lagos2018b} who showed that the higher spins in the most massive galaxies are due to AGN feedback being insufficient to quench star formation in the central galaxies of halos with masses $\gtrsim10^{14}\msun$ to the levels observed in brightest cluster galaxies. These galaxies end up having two times too much stellar mass and SFRs compared to observations. \ha\ galaxies at this mass range show more consistent values of \vse\ as compared to the observations (median \vse=0.11), whereas \ma\ has higher values and is more similar to \eap\ (median \vse=0.21) above $\logm>11.5$.

\begin{figure*}
\includegraphics[width=\linewidth]{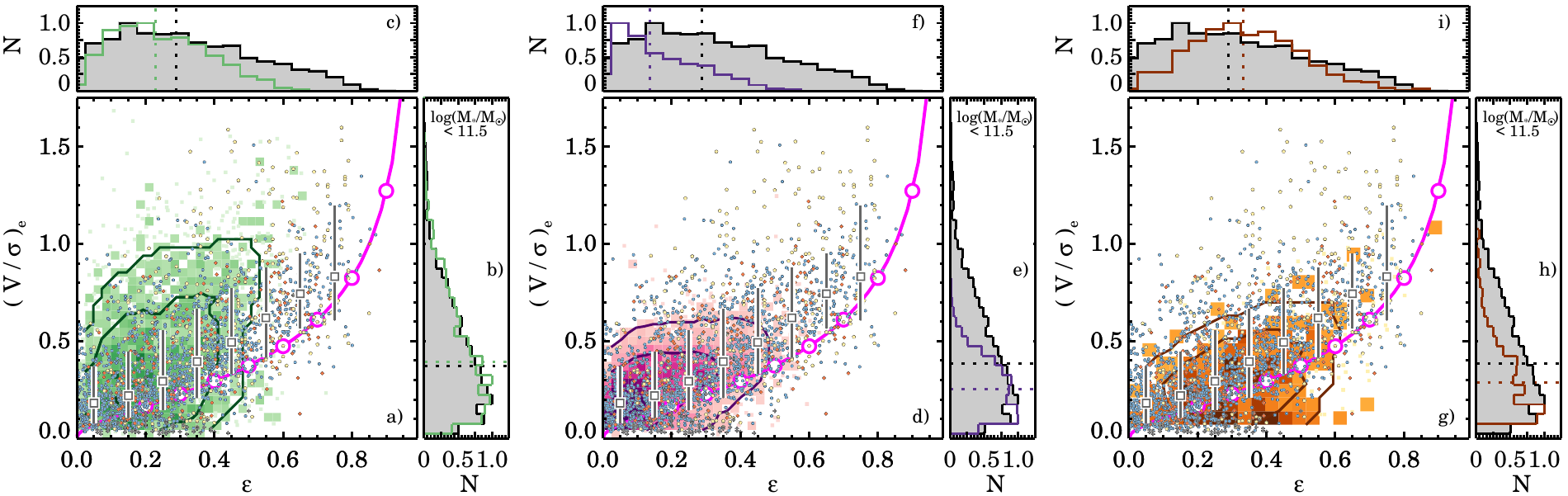}
\caption{Comparison of simulated and observational data in the \vs-$\epsilon$ diagram. Symbols are the same as in Figure \ref{fig:mass_size}, and the contours enclose 68 and 95 percent of the simulated data. The magenta line shows the theoretical prediction for an edge-on view of axisymmetric galaxies. On the line, from the bottom up, the open circles correspond to the locations of galaxies with different intrinsic ellipticities $\epsilon_{\rm{intr}}$=0.2-0.9 \citep[see][]{cappellari2007,emsellem2011}. Galaxies that are observed under different viewing angles follow a downward line from edge-on (magenta line) to face-on (towards zero ellipticity). As compared to observations, the shape of the distribution of \eap\ data is steeper, that is few simulated galaxies with $\ee>0.6$ and $\vse>0.5$ exist. For \ha\ the shape of the distribution appears to be correct, but both \ee\ and \vse\ are significantly lower as compared to observations. The \ma\ data closely matches the shape of the observed distribution but at fixed \ee\, galaxies with high \vse\ are missing.}
\label{fig:eps_vsigma}
\end{figure*}

\subsection{\vs-Ellipticity}
\label{subsec:eps_vsigma}

In Fig.~\ref{fig:eps_vsigma} we combine the ellipticity and \vse\ measurements, which allows us to study the dynamical properties of galaxies in relation to their intrinsic shape and inclination \citep[see e.g.,][]{binney2005, cappellari2007}. It is common practise to show tracks of different inclination and intrinsic ellipticity for oblate axisymetric rotators, however, to avoid over-crowding of data within the figures we only show model tracks for galaxies observed edge-on with varying intrinsic ellipticity (magenta-line). 

We previously found that the \eap\ galaxies are on average rounder, but with a similar distribution in \vse. From Fig.~\ref{fig:eps_vsigma}a-c we conclude that the \vse-\ee\ relation is steeper as compared to the observations; there are almost no simulated galaxies with $\ee>0.6$ and $\vse>0.5$. The \eap\ simulations recover the right ratio of velocity versus velocity dispersion, yet at fixed \vs\ galaxies are too round. Before, we also showed that the aperture velocity dispersions from the simulations were slightly too low (Fig.~\ref{fig:mass_sigma}). However, as the effective radii were also too large, the fact that the \vs\ values are correct here, does not directly imply that the average rotational velocities are too high. Whereas the shape of the \ha\ relation resembles the observational trend, we find that galaxies are both too round with too low \vs\ (Fig.~\ref{fig:eps_vsigma}d-f). Interestingly, we find a slow rotating horizontal band of galaxies with $\vse<0.15$ where the ellipticity extent almost out to $\ee=0.4$ that is very similar to the data from the MASSIVE Survey but to a lesser extent in the other surveys. This region of the $\vs-\epsilon$ diagram is typically occupied with galaxies that are triaxial in shape. In Fig.~\ref{fig:eps_vsigma}g-i we find that galaxies in the \ma\ closely match the observational data and the predictions for oblate axisymetric rotators (magenta line), while also recovering the region assigned to slow rotators. However, at fixed ellipticity, galaxies with high \vse\ values are absent, causing the overall distribution of \vse\ to be lower than observations.

\subsection{Age-Mass}
\label{subsec:mass_age}

For the second-last set of comparisons we switch to looking at stellar population properties, but limit ourselves to stellar population age. This is partly motivated by the recent SAMI Galaxy Survey result that intrinsic ellipticity and stellar age are correlated, independent of stellar mass or morphology \citep{vandesande2018}. These next two sections are therefore devoted to investigating whether this same relation exists within the simulations.

\begin{figure*}
\includegraphics[width=\linewidth]{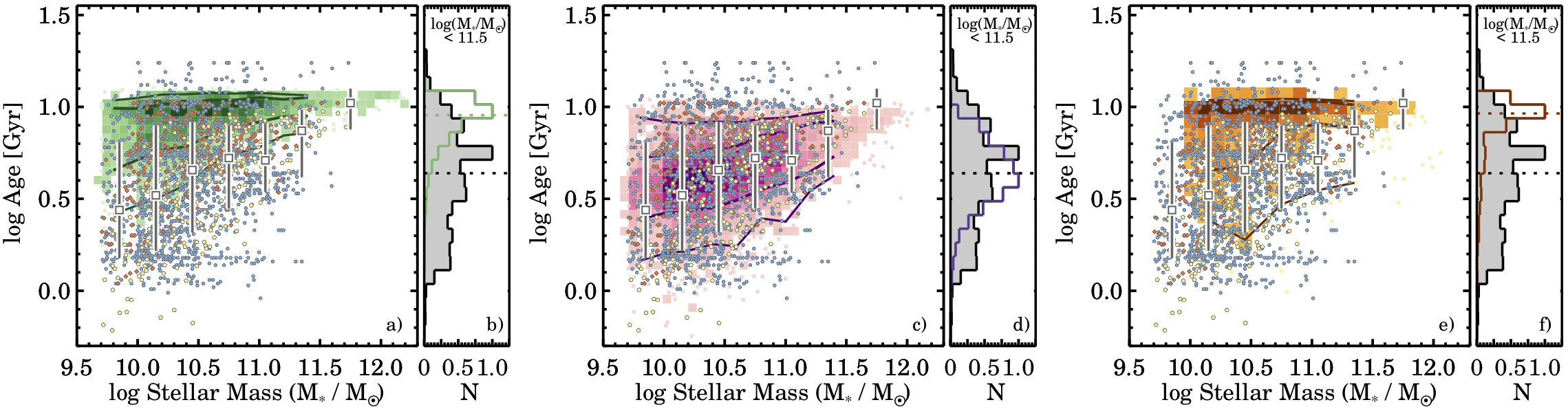}
\caption{Comparison of simulated and observational data in the age-mass diagram. Symbols are the same as in Figure \ref{fig:mass_size}, and the contours enclose 68 and 95 percent of the simulated data. For all three simulations we find a weak trend with stellar mass, where lower mass galaxies are younger. However, the median age for \eap\ and \ma\ is significantly higher as compared to observations. We find a large observed spread in age, whereas both \eap\ and \ma\ have a small spread in stellar population age. While \ha\ also shows a smaller spread in age as compared to observations, it is considerably more consistent as compared to the other two simulations.
}
\label{fig:mass_age}
\end{figure*}

In Fig.~\ref{fig:mass_age} we first present stellar population age as a function of stellar mass. The SAMI and CALIFA sample have a large spread in age $-0.2 < \log \rm{Age [Gyr]} < 1.2$, mostly due to the variety in morphologies from late-spirals to ellipticals. \at\ galaxies on average have older stellar populations, which is not surprising given that the sample consists of early-type galaxies only. 

In all three simulations there is a mild trend with stellar mass; on average galaxies are younger at lower stellar mass. In the \eap\ simulations we find no young galaxies with ages below $\log \rm{Age} < 0.4$ ($\sim2.5$ Gyr), and the spread of the age distribution is also significantly smaller. This could possibly be a consequence of \ea\ having a consistently higher fraction of passive low-mass satellite galaxies as compared to observations (Y. Bah\'e et al., in prep). The median observed log stellar age is 4.36 Gyr (2.5th percentile = 1.2 Gyr), whereas we find median age of 8.96 Gyr (2.5th percentile = 3.6 Gyr) in \eap. The median age of \ha\ is nearly identical to the observations, but similar to \eap\ we find a smaller spread in the age distribution (2.5th percentile log stellar age = 1.79 Gyr). The age distribution of \ma\ is very similar to \eap\, albeit with a few more young galaxies. However, the mismatch of \ma\ ages with the observations is large (median age 9.25 Gyr, lower 2.5th percentile = 2.35 Gyr).

There is a caveat that the observed luminosity-weighted SSP-equivalent ages are likely to be biased to young ages for all galaxies that have experienced recent star formation. While both observed and simulated ages are luminosity weighted, we expect the spectral modelling and choice of stellar population model to have a relatively large impact \cite[e.g.,][]{sanchez2016b}. We note that the observed stellar ages scatter well above the age of the Universe, which we regard as a caution to not over-interpret these results. In addition, no extinction effects were included in the $r$-band luminosity weighting in the simulations.

\subsection{Intrinsic Ellipticity - Age}
\label{subsec:eps_intr_age}

For the final comparison, we explore the relation between the intrinsic ellipticity or edge-on ellipticity of galaxies and their mean stellar population age. In \citet{vandesande2018}, we showed that the characteristic stellar age follows the intrinsic ellipticity of galaxies remarkably well. As the shape and kinematic properties of the galaxies change from dynamically cold, and intrinsically flat into dynamically-hotter, pressure-supported, thicker, oblate spheroids, we find that the mean age of the stellar population increases. This trend is still observed when galaxies are separated in bins of low and high-stellar mass, or early-type and late-type. We note that this trend is more than a different empirical view of the relation between stellar population properties and bulge fraction \citep[e.g.,][]{cappellari2016}; the relation between intrinsic ellipticity and age is also found for low mass galaxies ($\log M_{*}/\msun<10.25$), where the fraction of galaxies with a classical, dispersion-dominated bulge is low \citep[$<30$ percent;][]{vandesande2018}.

We repeat the analysis of \citet{vandesande2018} with the full sample, with the exception of MASSIVE galaxies for which we do not have stellar age measurements. For details on the technique to derive intrinsic ellipticities we refer to their work, but we highlight the main assumptions here. First, we assume that all galaxies are axisymmetric, and mildly anisotropic: $\beta_z = 0.6\pm 0.1\times \epsintr$  \citep{cappellari2007} where $\beta_z$ is the anisotropy and $\epsintr$ is the intrinsic ellipticity. Secondly, any galaxy that falls below the magenta line in Fig.~\ref{fig:eps_vsigma} is excluded from the sample because they are outside the model range. These are typically the galaxies with the oldest stellar populations. Third, near-perfectly round galaxies ($\ee<0.025$) are also excluded, because the model predictions are highly degenerate and the relative measurement uncertainties on \ee\ are large. 

In the simulations we do not have to rely on any of the de-projection assumptions for deriving intrinsic ellipticities. Instead, we can use the edge-on projection of all galaxies and measure the intrinsic ellipticity from those. Here, the edge-on projection angle is computed using the total stellar angular momentum of all stellar particles in the galaxy. Similar to the observed sample, we exclude all \eap, \ha, and \ma\ galaxies below the magenta line in Fig.~\ref{fig:eps_vsigma} so that both simulated and observed samples are selected in the same way.

We note that we could have derived \epsintr\ using an identical method that is used for the observations. However, we already concluded in Section \ref{subsec:eps_vsigma} that the simulated data occupy different regions of the $\vs-\epsilon$ diagram as compared to observations. The observed intrinsic ellipticity distribution derived from \vse\ and \ee\ is similar to intrinsic ellipticity distributions derived from inverting the distributions of apparent ellipticities and kinematic misalignments \citep{weijmans2014,foster2017}. However, as the simulated data does not overlap with the observations presented here in either \vs\ or $\epsilon$, using the same approach will not result in accurate intrinsic ellipticity. Therefore, we use the edge-on projection in the simulations to estimate intrinsic ellipticities.

\begin{figure*}
\includegraphics[width=\linewidth]{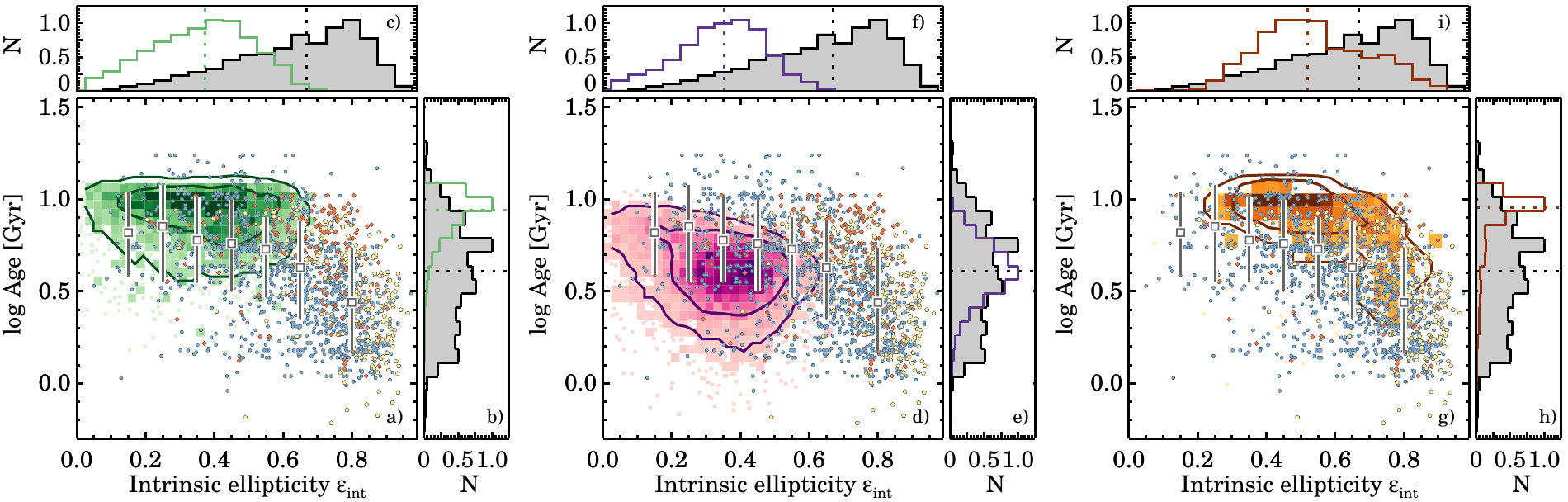}
\caption{Comparison of simulated and observational data in the age-intrinsic ellipticity diagram. Symbols are the same as in Figure \ref{fig:mass_size}. The intrinsic ellipticities from the simulations are derived from the edge-on projection of galaxies, whereas the observational intrinsic ellipticities have been derived using theoretical model predictions of axisymmetric oblate rotators. The trend where younger galaxies are also more flattened is poorly recovered in \eap, slightly more pronounced in \ha, but clearly detected in \ma. However, at fixed intrinsic ellipticity, all simulations show offsets in age. In both \eap\ and \ha\ there is a deficit of flattened galaxies with $\epsintr>0.7$, while \ma\ galaxies extend to $\epsintr\sim0.7$.}
\label{fig:eps_intr_age}
\end{figure*}

In Fig.~\ref{fig:eps_intr_age} we show the relation between age and intrinsic ellipticity from observations and simulations. There is a good agreement between the SAMI, \at\, and the CALIFA trends, but as expected the CALIFA sample contains more galaxies with intrinsic ellipticity that are highly flattened. We find that the shape of the \eap\ age-intrinsic ellipticity distribution is mildly consistent with the observations, although the \eap\ distribution is offset to older ages. The distribution from the \eap\ simulations is also not as extended as the observed relation (median \epsintr\ of 0.37 and 0.67 for \eap\ and observation, respectively), and the scatter in age is also considerably smaller. The regime where the observational ages significantly start to decrease at $\epsintr>0.6$ is not covered by \eap\ thus we cannot conclude whether there is a trend between age and intrinsic ellipticity. \ha\ galaxies show a qualitatively consistent trend with the observations where younger galaxies are more flattened (lower \epsintr). Yet, while \ha\ has relatively young stellar ages, the galaxies are surprisingly round (median \epsintr\ of 0.35). We do find a clear trend between age and \epsintr\ in the \ma\ simulations, albeit with an offset towards older ages. \ma\ galaxies extend significantly beyond both \eap\ and \ha\ in terms of intrinsic ellipticity. Flattened galaxies with high intrinsic ellipticity ($\epsintr>0.7$) that are abundant in the observed samples are only seen in \ma, but the overall \epsintr\ distribution in \ma\ is offset to lower values (median \epsintr\ of 0.53).

Naively, some of the discrepancies can be understood by the simple lack of high ellipticity galaxies or from the lack of galaxies with young stellar ages. In the first scenario one would expect that future simulations in which high ellipticity galaxies exist, would naturally extend the relations seen here for \eap\ and \ha\ to the high ellipticity population. However, this may not be so simple, as the changes needed to reproduce the vertical structure of disks, and therefore the high-ellipticity galaxies, are major (i.e., higher resolution, more realistic ISM models, and a corresponding stellar feedback that acts on smaller scales). Thus, it is not guaranteed that the properties of the rounder, massive ellipticals and lenticulars - that the current generation of simulations reproduce well - will still be reproduced by the next generation of simulations.

\section{Previous Comparison Studies}
\label{sec:previous}

We have now compared seven different structural, dynamical, and stellar population parameters from observations and mass-matched simulations at $z<0.1$. Before summarising our findings and discussing their implications, here we first briefly review the literature from the last decade and then review previous work that used the same simulations in comparison to observations.

\subsection{Results from Simulations in the Last Decade}
\label{subsec:previous_decade}

The last decade has seen significant progress in the area of cosmological hydrodynamical simulations, with the first large cosmological volumes, with high enough resolution to study the internal structure of galaxies being possible \citep[e.g.,][]{devriendt2010,dubois2014,vogelsberger2014,schaye2015,Pillepich18}. These simulations have been designed to overcome the problem of catastrophic loss of angular momentum and over-cooling problem that was observed in simulations prior to 2010 (\citealt{Steinmetz99}; \citealt{Navarro00}). The catastrophic loss of angular momentum problem refers generally to the issue of simulated galaxies being too compact and having too low specific angular momentum, which affected Smoothed Particle Hydrodynamics (SPH) codes mostly \citep[e.g., GADGET,][]{springel2001}. This  problem that was mainly solved by implementing angular momentum conserving SPH schemes. On the other hand, Adaptive Mesh Refinement (AMR) codes \citep[e.g., RAMSES,][]{teyssier2002} did not suffer from the same angular momentum problem, but only from the over-cooling problem \citep{lackner2012,Dubois13}. In AMR, massive galaxies tended to be too large and have prominent disks, in contradiction with the observations of massive galaxies. Thus, in general both SPH and AMR codes generated galaxies were too massive compared to observations due to the efficient cooling and star formation, and in addition too compact in the case of SPH. The angular momentum catastrophe and over-cooling problem were solved by a combination of improved spatial resolution and numerical techniques of the simulations and the inclusion of efficient feedback 
\citep[e.g.,][]{Kaufmann07,Zavala08,Governato10,Guedes11,DeFelippis17}. 

Feedback was necessary to prevent large amounts of gas from cooling onto the galaxies and therefore fuelling star formation, to avoid the build-up of stellar mass in the centres of galaxies, and for removing low angular momentum gas from galaxies. For example, \citet{remus2017} showed that the properties of early-type galaxies can only be reproduced with AGN. Without AGN, early-type galaxies are too compact with dark matter fractions that are too low. Because of the catastrophic angular momentum problem, the comparison of kinematic measurements of galaxies and simulations was mostly done in a qualitative manner, or in idealised simulations tackling specific physical processes (such as galaxy mergers; e.g., \citealt{cox2006}; \citealt{taranu2013}; \citealt{naab2014}).  

There has been a considerable amount of work over the last years comparing the specific angular momentum of galaxies with observations of late- and early-type galaxies in the local Universe \citep[e.g.,][]{Pedrosa15,Teklu15,Genel15,Zavala16,lagos2017,wang18}, whereas other studies compare \lr, the proxy for the stellar spin parameter, from observations in early-type galaxies \citep{emsellem2007,emsellem2011} with simulations \citep[e.g.,][]{naab2014,penoyre2017,choi2017,lagos2018b,schulze2018}. Some comparisons between observations and simulations have also been presented for galaxies at high redshift \citep{Swinbank17}. All this work has shown that simulations are now getting close to reproducing observations, with disks typically having high spin and high specific angular momentum, whereas bulges have $\approx 5-8$ times less specific angular momentum at fixed stellar mass.

\subsection{Comparing Observations and Simulations with EAGLE, HORIZON-AGN, and MAGNETICUM}
\label{subsec:previous}

We now review previous work that used \ea, \ha, or \ma, for a comparison to observations. The size-mass relation for \ea\ is presented in \citet{schaye2015}, where the simulations are compared to SDSS results from \citet{shen2003} for disk-like galaxies, i.e., where \ser\ $n<2.5$. The agreement they find is excellent, albeit by construction, as the models that produced galaxies that were too small were rejected \citep{crain2015}. This calibration was derived by using the size-mass relation of star-forming galaxies. The difference between their models and observations is $\lesssim0.1$ dex over the full stellar mass range. The size-mass relation in \ea\ is also studied in \citet{furlong2017}, at different redshifts and separating star-forming and passive galaxies. They too find a good agreement to the \citet{shen2003} results. We note that \citet{furlong2017} apply a constant factor of 1.4 to convert the \citet{shen2003} circularised radii to semi-major axis values of both star forming \emph{and} passive galaxies. They find systematically larger radii ($\sim0.2$ dex) as compared to \citet{schaye2015}, but this mismatch in radii is attributed to different measuring techniques. Nonetheless, \citet{furlong2017} conclude that the \ea\ size-mass relation shows an excellent agreement with observations.

Instead of using the \citet{shen2003} size-mass relations, \citet{lange2016} use reprocessed SDSS imaging as part of the GAMA Survey as their observational reference. They find an excellent agreement of spheroidal galaxies down to $\logm \sim10$, but below this mass \ea\ galaxies are too large. Disk galaxy sizes are found to be larger in the GAMA data. This is partly because disk-only components are derived for GAMA data (from bulge/disk decomposition), which are then compared with global sizes from \ea. 

In general, the main conclusions from these papers are slightly different to the results presented here; we find that the size-mass from \ea\ is offset to larger sizes as compared to the observational data. An important caveat here is that previous work used observational data with optical luminosity effective radii, while for \ea\ stellar mass half-mass radii were used. This can lead to a systematic offsets when one compares half-light versus half-mass radii, as half-light radii tend to be larger than the stellar mass-based ones (see Appendix \ref{sec:app_lum_mass}). Thus, a significant improvement for future simulations is to make more realistic comparisons with the measured sizes of galaxies by taking into account the way they have been measured.

The \ha\ size-mass relation is compared to observational results in \citet{dubois2016}. Their conclusion is that the sizes at $z = 0.25$ are in good agreement with observational results from \citet{vanderwel2014}, who use deep \textit{Hubble Space Telescope} data to investigate the size-mass evolution of galaxies. For early-type galaxies, the \ha\ size-mass relation is higher in size by $\sim 0.2$ dex as compared to late-type galaxies at a stellar mass of $\logm<11$. This is inconsistent with the observational result from \citet{vanderwel2014}, and with the results presented here. While we do not split the sample into late- and early-type, we find that \ha\ galaxies are on average too large at all stellar masses as compared at $z<0.1$, which appears to be in contrast with the results from \citet{dubois2016}. The most likely reason for late-type and early-type to have a lower size-mass relation in \citet{dubois2016} is because the effective radii are mass-weighted quantities and not $r$-band luminosity weighted. As luminosity weighting typically increases the sizes of late-type galaxies more than early-type galaxies, this could explain the apparent inconsistency.

 \citet{remus2017} compare the \ma\ size-mass relation for early-type galaxies to results from SDSS \citep{shen2003} and GAMA \citep{baldry2012} and find that AGN feedback is essential to establish the observed size-mass relations. \cite{schulze2018} also investigate the \ma\ size-mass relation, but now for both early-type and late-type galaxies to observations from \at, CALIFA, SLUGGS \citep{forbes2017}, and GAMA \citep{baldry2012,lange2015}. Note that both \citet{remus2017} and \citet{schulze2018} use mass-weighted radii from \ma\ in the comparison but luminosity-weighted sizes from observations. \citet{schulze2018} conclude that the simulated \ma\ sizes are consistent with GAMA observations, but that they are larger than the sizes from IFS observations. This is contradictory to our results. When using luminosity-weighted radii and a mass-matched sample, we find that at fixed stellar mass, \ma\ sizes are overestimated. Similar as for the previous \ea\ comparisons, this can be attributed to the fact that luminosity-weighted sizes on average are larger than mass-weighted sizes (see Appendix \ref{sec:app_lum_mass}).

The dynamical properties of simulated galaxies in \ea, \ha, and \ma\ have been studied in several papers. \citet{chisari2015} use \ha\ and present stellar rotational properties of galaxies as a function of stellar mass. As their definition of \vs\ is different from the commonly used definition used in observations, their results were not compared to observations. Instead of measuring velocity and velocity dispersion from an LOSVD, $V$ and $\sigma$ are derived from decomposing the spin properties, i.e., the total angular momentum, of individual star particles. Thus \vs\ is an average derived property using all individual star particles. The trend in \vs\ with stellar mass is similar as found here, but due to the different definition of \vs\, we can only make a qualitative comparison.

Using the same data, \citet{dubois2016} investigate the impact of AGN on the \vs\ measurements and find that without AGN feedback their simulation produces too many fast-rotating galaxies above a stellar mass of \logm>10.5. \citet{choi2018} also use the \ha\ simulations with an approach similar to ours, i.e., they extract galaxy parameters from the simulations with measuring techniques that closely match observations. They use the spin parameter proxy \lr, which is nearly identical to \vs, and investigate the evolution of early-type galaxies. Similar to our results, they find that the majority of galaxies in \ha\ have low spin ($\lre \lesssim 0.3$) and are rounder ($\epsilon<0.2$) as compared to observed samples. Nonetheless, they successfully recover the observed trend that the most massive galaxies have the lowest spin.

\ma\ early-type galaxies are compared to SAMI, \at, CALIFA, and SLUGGS in the $\lre-\epsilon$ plane in \citet{schulze2018}. Their analysis also includes a cumulative distribution comparison. They conclude that 70 percent of their galaxies are classified as fast-rotators and 30 percent as slow-rotators using the \citet{emsellem2011} criterion. While the number of fast-rotators is lower as compared to \at\ \citep[86 percent,][]{emsellem2011}, these numbers are qualitatively similar to what we find here. When comparing \vse\ from \ma\ to the full observed sample, without a cut in morphology, we find that the \ma\ galaxies have moderately lower \vse\ values.

The dynamical properties of galaxies in the \ea\ simulations are studied by \cite{lagos2017,lagos2018a,lagos2018b}. They find an excellent match between the specific angular momentum of galaxies ($j_\star$) in \ea, \at, and SAMI as presented in \citet{cortese2016}. Using both \ea\ and \hy, \citet{lagos2018b} focus on the impact of galaxy mass and environment on the spin-down of galaxies using the spin parameter proxy. When comparing observations and simulations in the \lre-\ee\ plane, they find similar results as presented here. Because the focus in \citet{lagos2018b} was placed on massive, slowly rotating galaxies, our comparison here extends to significantly higher spin parameter values and high ellipticity galaxies.

The comparisons presented here of ellipticity, velocity dispersions, and age, have not been analysed in this much detail before. Thus we have the ability to unveil new and important shortcomings of the simulations. Ellipticity has been explored in combination with \vs\ and \lr\ \citep[e.g.,][]{naab2014,penoyre2017,choi2018,lagos2018b, schulze2018}, but here we have taken an additional step of comparing ellipticity distributions using mass-matched samples. The offset towards thicker disks in \ea\ was also noticed by \citet{trayford2017} who found that the optical extinction is much smaller than observed for edge-on galaxies, but consistent for face-on objects. Their interpretation was that \ea\ disks are too thick. Stellar ages are a key aspect where we can advance our understanding of galaxy formation by informing the upcoming simulations of areas in which more development is needed. There have been several papers looking at stellar population properties in simulations, in particular colour \citep[e.g.,][]{trayford2015,correa2017}. These results show that the colour of simulated galaxies match the observed loci of the blue-cloud and red-sequence, though not the width of the colour distributions. This agreement may be inconsistent with our findings that the average age of low-redshift \ea\ galaxies is too high. However, colours and ages, though correlated, are not equivalent to each other and therefore the discrepancies here may come from additional effects, such as metallicity, dust, or systematic differences in the way ages and environments are measured in the simulation versus observations.

Our work pushes the boundaries of the comparisons done to date in the sense that we explore a very wide range of galaxy properties. Simultaneously, we take special care in the way we are calculating all galaxy properties to mimic the observations. The detail in which we compare to simulations, both qualitative and quantitative, and using both the median and the spread of the distributions, is also unprecedented. This has been largely feasible because simulations are now able to predict galaxy properties that match better with observations than ever before.

\section{Summary and Discussion}
\label{sec:summary_discussion}

This study presents a detailed comparison of multiple galaxy properties from simulations and IFS observations using as close to identical measurement techniques as possible. We compare results from the \ea, \hy, \ha, and \ma\ simulation projects to observational results from the SAMI Galaxy Survey, CALIFA, \at\, and MASSIVE Surveys. The main goal of this work is to take advantage of the recent wealth of 2D IFS data and compare these to cosmological simulations, with the aim of identifying key areas of success and tension. 
By investigating whether simulations simultaneously reproduce the structural, resolved dynamical and stellar population properties of galaxies, any area of tension will inform us where the simulations require revision. Furthermore, the advantage of using more than one cosmological hydrodynamical simulation allows us to compare the results from different simulations with each other. 

Our measurements from the simulations closely match the observational techniques. In particular for the dynamical measurements, this involves constructing line-of-sight velocity distributions, from which the velocity and velocity dispersion were extracted. We combine four different IFS surveys to create a more homogeneous observational sample where individual survey biases are lessened. A mass-matching technique is used to remove the difference between the observed and simulated stellar mass function. With this method, we sample the same fraction of galaxies from both observations and simulations as a function of stellar mass.

For the \ea\ and \hy\ simulations, here combined and referred to as \eap, we find a good match with the observations in the size-mass plane, although there is an offset to larger effective radii. The trend of observed ellipticity with stellar mass is in good agreement, but the thinnest/flattest galaxies are not present in the simulations. At fixed stellar mass, the aperture velocity dispersions are too low. Yet when we combine size and velocity dispersions into dynamical masses, and compare these to stellar masses, there is an excellent agreement with observations. The ratio of velocity and velocity dispersion \vs\ match remarkably well with observations. However, not all regions of the \vs-\ee\ space that are covered by observations are matched by \eap. The median stellar age of galaxies in \ea\ is higher than in observations, and galaxies with young stellar populations are absent. We do not recover the recently discovered observational trend between intrinsic ellipticity and age in \eap. This is can be attributed to the fact that both young and intrinsically flat galaxies are missing, thus we cannot conclude whether a relation between intrinsic ellipticity and age exists in \eap. We find the best match between observations and the \eap\ simulations in the low to mid stellar mass regime below $\logm<11$. At higher stellar masses, \eap\ galaxies tend to retain or build-up too much spin. 

The \ha\ simulations provide a qualitatively decent match to the observations, i.e., many of the trends with stellar mass are recovered. Quantitatively, there are several parameters where the median is significantly lower or higher, or where the range in \ha\ is smaller as compared to observations. While the size-mass relation has the right slope and scatter, we find that simulated galaxies are offset towards larger size. As a function of stellar mass, galaxies in \ha\ become rounder, but the median ellipticity is too low at all stellar masses, except above $\logm>11.5$. Aperture velocity dispersions are too low as compared to observations, and the slope of the $\sigma$-mass relation is slightly too shallow. When comparing dynamical masses to stellar masses we find trends similar to observations, but with less scatter at fixed stellar mass. At the highest stellar mass galaxies ($\logm>11.3$), we find a good match between the ratio of velocity and velocity dispersion \vs. For galaxies with stellar mass $\logm<11$, fast-rotating galaxies with $\vse>0.75$ appear to be absent. \ha\ recovers the \vse-\ee\ relation across the range that it resolves, but a large fraction of the \vse-\ee\ space covered by observations is not occupied by \ha\ galaxies. We find an excellent match in the median age of galaxies, but the spread in age is smaller in the simulations. We recover a weak trend between intrinsic ellipticity and age in \ha, but for a given stellar age, we find that \ha\ have intrinsic ellipticity that are too low as compared to observations. The best match between observations and simulation is observed at high-stellar masses ($\logm>11.3$). At all other masses galaxies are typically too big, too round, with values of \vse\ that are too low. 

For the \ma\ simulations, we find a good qualitative and quantitative match for many of the trends with stellar mass and (intrinsic) ellipticity. A complication in all the comparisons of the \ma\ data with observations is the relatively small volume of the cosmological box. Nonetheless, the size-mass relation is well reproduced both the slope and the centre of the distribution. Flat observed galaxies (high $\epsilon$) are produced in \ma, but there is a deficit of extremely round galaxies. Galaxies decrease in $\epsilon$ with increasing stellar mass, but the onset of this decline appears to at lower stellar mass as compared to observations. Velocity dispersions are too low at all stellar masses, but dynamical mass estimates agree well with observations. \vs\ measurements in \ma\ are moderately too low at fixed stellar mass and the declining trend of \vs\ with stellar mass starts at lower stellar mass as compared to observations. We find a good match in \vse-\ee\ space, i.e., the simulated galaxies follow the same trend as the observations, whilst also producing slow-rotating galaxies. However, at fixed ellipticity high \vse\ galaxies values are missing, and the overall distribution of \vse\ is lower as compared to observations, i.e., the relative number of fast spinning galaxies is still too low. The derived mean stellar age in \ma\ is always higher than the observed age distribution. We find a clear confirmation of the relation between age and intrinsic ellipticity in \ma\, albeit with an offset towards older ages. We find the best match between observations and the \ma\ simulations in the mid-to-high stellar mass regime above $\logm>10.75$. While there is not a strong trend between how well the simulations match the observations and stellar mass, it is clear that higher particle resolution is needed to test \ma\ predictions at low stellar mass. 

The implications of these results can be separated into two stellar mass regimes. Without comparing simulations to each other, we find that the \eap\ simulations reproduce several of the relations at low and intermediate stellar masses ($\logm<11$) better than at high stellar mass. The \ha\ simulations reproduce the properties of the most massive galaxies ($\logm>11.3$) considerably better as compared to low-mass \ha\ galaxies. Similarly, \ma\ also appears to perform better at high-stellar mass than low stellar mass. 

As shown in \citet{dubois2016}, the biggest impact of their two-mode AGN-feedback in \ha\ is at the massive end of the galaxy mass function. In \ea\ the treatment of AGN feedback is implemented in a different way, but given that the most massive galaxies typically retain too much spin, this suggests that the feedback is not effective enough. This agrees with the fact that \eap\ galaxies in clusters are about a factor of $2$ too massive and have star formation rates that are higher than observed in brightest cluster galaxies \citep{bahe2017,barnes2017}. In \citet{lagos2018b} this was seen as the fraction of slow rotators being too low at the highest stellar masses compared to observations. In \ha, at $\logm<11$, where AGN feedback is less dominant, the simulations do not provide a good quantitative match to the observations.

Cosmological hydrodynamical simulations are rich tools to disentangle the relative importance of the physical processes involved in galaxy formation and evolution. The comparison with IFS measurements has shown that some fundamental parameters that describe a galaxy are not well reproduced in simulations. Moreover, the areas of discrepancy and agreement are different in the various simulations. Thus, there are a number of caveats which concern the validity of using these mismatching parameters to study the evolution of galaxies. Our comparison has highlighted several areas where the simulations might be improved, among which are the limiting particle resolution and the ISM model.

Based on our results, we consider the next major challenge for cosmological hydrodynamical simulations of galaxy formation is to produce Milky Way analogue galaxies that have both dynamically \emph{and} chemically distinct thick and thin disks \citep[e.g.,][]{navarro2018}. In this task, it will be essential to reproduce realistic vertical structure in galaxy disks. The next major observational step will be to gather larger statistical samples of resolved stellar kinematic measurements \citep[e.g., Hector,][]{bryant2016} and to go beyond $z\sim0.1$. With ever growing IFS surveys on galaxies near and far, the next generation of simulations will have even better samples to which to calibrate.



\section*{Acknowledgements}

We thank the anonymous referee for the very constructive comments which improved the quality of the paper. 

We thank Joop Schaye, Klaus Dolag, Matthieu Schaller, and Tom Theuns for useful discussions and constructive comments. 

The SAMI Galaxy Survey is based on observations made at the Anglo-Australian Telescope. The Sydney-AAO Multi-object Integral-field spectrograph (SAMI) was developed jointly by the University of Sydney and the Australian Astronomical Observatory, and funded by ARC grants FF0776384 (Bland-Hawthorn) and LE130100198. JvdS is funded under Bland-Hawthorn's ARC Laureate Fellowship (FL140100278). CL has received funding from a Discovery Early Career Researcher Award (DE150100618) and by the ARC Centre of Excellence for All Sky Astrophysics in 3 Dimensions (ASTRO 3D), through project number CE170100013. RSR and FS acknowledge support from the DAAD through funds from the Federal Ministry of Education and Research (BMBF). SB acknowledges the funding support from the Australian Research Council through a Future Fellowship (FT140101166). JJB acknowledges support of an Australian Research Council Future Fellowship (FT180100231). LC is the recipient of an Australian Research Council Future Fellowship (FT180100066) funded by the Australian Government. NS acknowledges support of a University of Sydney Postdoctoral Research Fellowship. Support for AMM is provided by NASA through Hubble Fellowship grant \#HST-HF2-51377 awarded by the Space Telescope Science Institute, which is operated by the Association of Universities for Research in Astronomy, Inc., for NASA, under contract NAS5-26555.

The SAMI input catalogue is based on data taken from the Sloan Digital Sky Survey, the GAMA Survey and the VST ATLAS Survey. The SAMI Galaxy Survey is supported by the Australian Research Council Centre of Excellence for All Sky Astrophysics in 3 Dimensions (ASTRO 3D), through project number CE170100013, the Australian Research Council Centre of Excellence for All-sky Astrophysics (CAASTRO), through project number CE110001020, and other participating institutions. The SAMI Galaxy Survey website is http://sami-survey.org/ .

The \ma\ Pathfinder simulations were partially performed at the Leibniz-Rechenzentrum with CPU time assigned to the Project 'pr86re'. This work was supported by the DFG Cluster of Excellence 'Origin and Structure of the Universe'. We are especially grateful for the support by M. Petkova through the Computational Center for Particle and Astrophysics (C2PAP).

This study uses data provided by the Calar Alto Legacy Integral Field Area (CALIFA) survey (http://califa.caha.es/). Based on observations collected at the Centro Astron\'omico Hispano Alem\'an (CAHA) at Calar Alto, operated jointly by the Max-Planck-Institut f\"ur Astronomie and the Instituto de Astrof\'isica de Andaluc\'ia (CSIC).



\bibliographystyle{mnras}
\bibliography{jvds_sami_sim} 

~\\
{\it \footnotesize \noindent$^{1}${Sydney Institute for Astronomy, School of Physics, A28, The University of Sydney, NSW, 2006, Australia}\\
$^{2}${ARC Centre of Excellence for All Sky Astrophysics in 3 Dimensions (ASTRO 3D), Australia}\\
$^{3}${International Centre for Radio Astronomy Research, The University of Western Australia, 35 Stirling Highway, Crawley WA 6009, Australia}\\
$^{4}$ Universit\"ats-Sternwarte M\"unchen, Scheinerstr. 1, D-81679 M\"unchen, Germany \\
$^{5}$ Max Planck Institute for Extraterrestrial Physics, Giessenbachstra{\ss}e 1, D-85748 Garching, Germany \\
$^{6}$ Leiden Observatory, Leiden University, PO Box 9513, NL-2300 RA Leiden, the Netherlands \\
$^{7}${School of Physics, University of New South Wales, NSW 2052, Australia}\\
$^{8}${Australian Astronomical Optics, AAO-USydney, School of Physics, University of Sydney, NSW 2006, Australia}\\
$^{9}${University of Oxford, Astrophysics, Keble Road, Oxford OX1 3RH, UK}\\
$^{10}${CNRS and UPMC Univ. Paris 06, UMR 7095, Institut d'Astrophysique de Paris, 98bis boulevard Arago, 75014 Paris, France} \\
$^{11}${Australian Astronomical Optics, Faculty of Science and Engineering, Macquarie University, 105 Delhi Rd, North Ryde, NSW 2113, Australia}\\
$^{12}${Atlassian, 341 George St, Sydney, NSW 2000, Australia}\\
$^{13}${Ritter Astrophysical Research Center University of Toledo Toledo, OH 43606, USA}\\
$^{14}${Research School of Astronomy and Astrophysics, Australian National University, Canberra ACT 2611, Australia}\\
$^{15}${Hubble Fellow}\\
$^{16}${Korea Institute for Advanced Study (KIAS), 85 Hoegiro, Dongdaemun-gu, Seoul, 02455, Republic of Korea}\\
$^{17}${SOFIA Science Center, USRA, NASA Ames Research Center, Building N232, M/S 232-12, P.O. Box 1, Moffett Field, CA 94035-0001, USA}\\
$^{18}${Instituto de Astronom\'{i}a, Universidad Nacional Aut\'{o}noma de M\'{e}xico, A. P. 70-264, C.P. 04510 M\'{e}xico, D.F. Mexico}\\
$^{19}$Centre for Astrophysics and Supercomputing, Swinburne University of Technology, PO Box 218, Hawthorn, VIC 3122, Australia \\}

\appendix

\begin{figure*}
\includegraphics[width=0.85\linewidth]{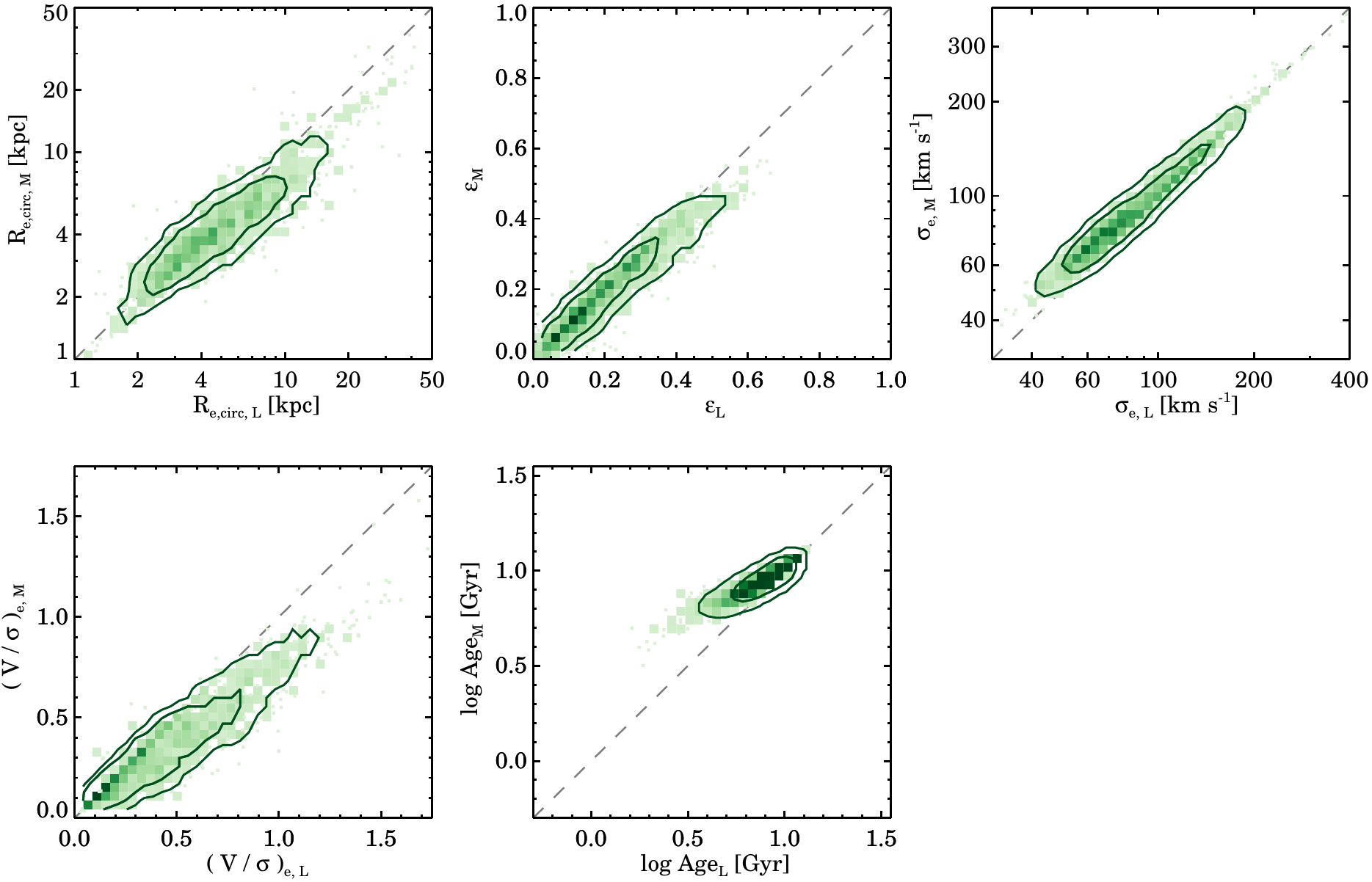}
\caption{Comparison of simulated luminosity and mass-weighted quantities in \ea. The assumption of using luminosity-weighted measurements rather than mass-weighting, results in larger effective radii, higher or more flattened ellipticities, lower velocity dispersions, higher \vse, and younger mean ages.
}
\label{fig:lum_mass_weighting_ea}
\end{figure*}

\begin{figure*}
\includegraphics[width=0.85\linewidth]{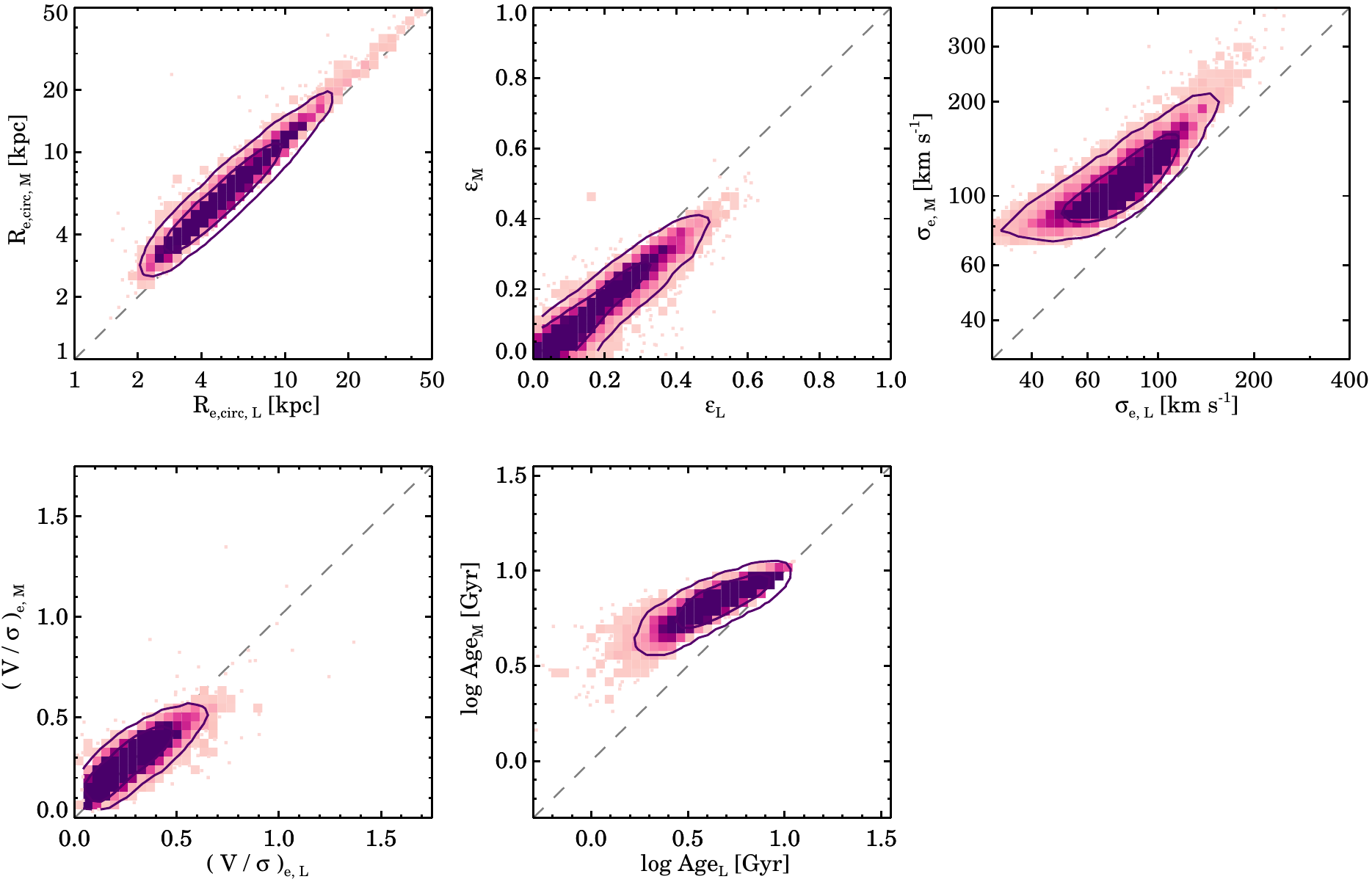}
\caption{Comparison of simulated luminosity and mass-weighted quantities in \ha. Similar to \ea, we find that with luminosity-weighted ellipticities are higher, velocity dispersions are considerably lower, \vse\ values are higher, and ages are younger. However, the luminosity-weighted effective radii are smaller than the mass-weighted radii, in contrast with observational results.
}
\label{fig:lum_mass_weighting_ha}
\end{figure*}

\begin{figure*}
\includegraphics[width=0.85\linewidth]{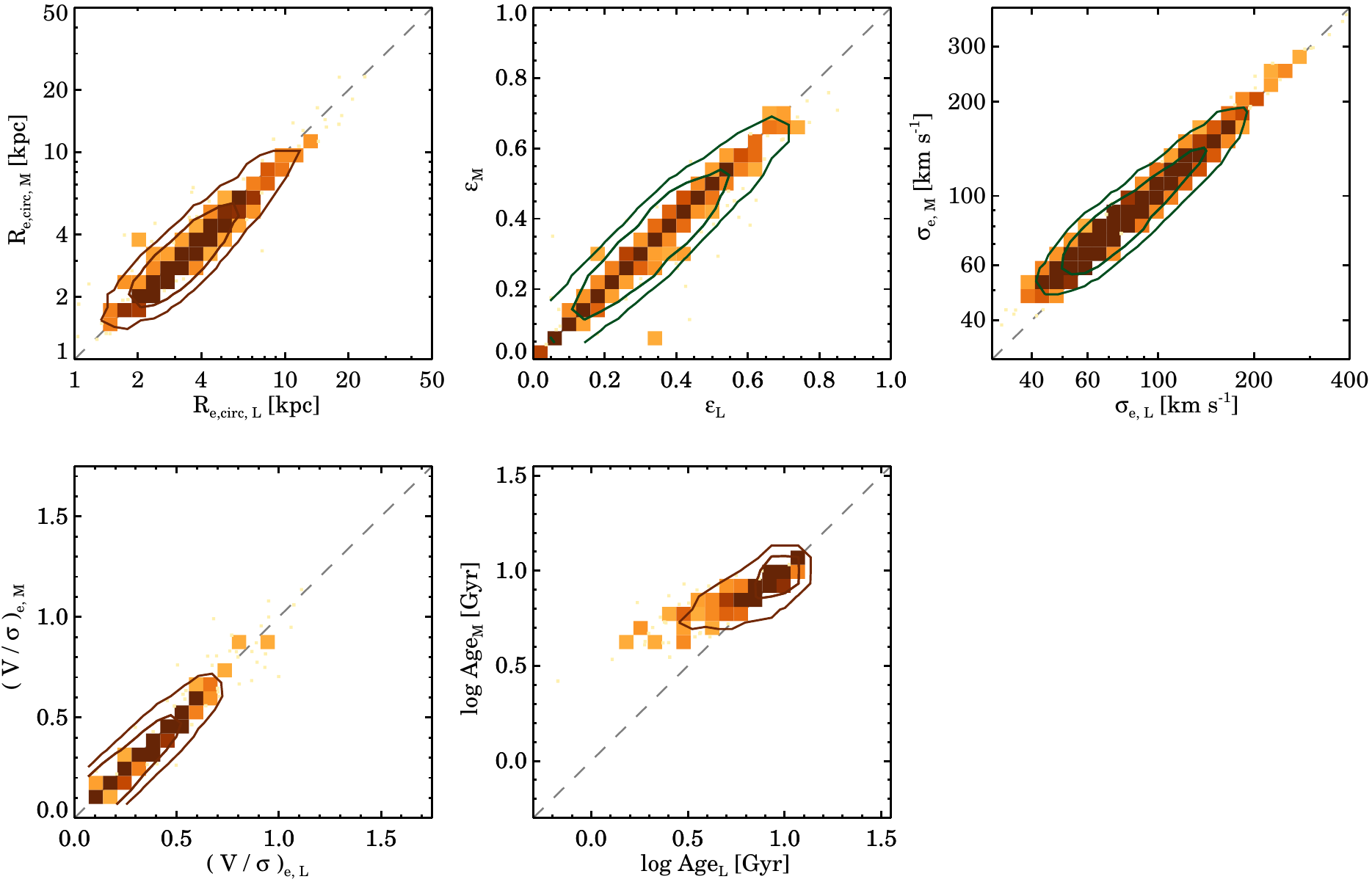}
\caption{Comparison of simulated luminosity and mass-weighted quantities in \ma. Similar to \ea, we find that with luminosity-weighted sizes and ellipticities are higher, velocity dispersions are considerably lower, \vse\ values are slightly higher, and ages are younger.
}
\label{fig:lum_mass_weighting_ma}
\end{figure*}

\section{Luminosity and Mass Weighted Quantities}
\label{sec:app_lum_mass}
To provide the best-match between the simulations and IFS observations, in this paper all mock-observations are luminosity weighted. In this appendix, we study the impact of using $r$-band luminosity-weighted versus stellar mass-weighted quantities. In Fig.~\ref{fig:lum_mass_weighting_ea}-\ref{fig:lum_mass_weighting_ha} we compare these two different weighting approaches for the five core measurements used in this work: effective radius \re, ellipticity \ee, velocity dispersion \se, \vse, and the mean stellar population age.

For the \ea\ simulations (Fig.~\ref{fig:lum_mass_weighting_ea}), we find trends consistent with observations. Mass-weighted effective radii are smaller than mass-weighted radii, as expected for galaxies with negative colour gradients where the cores are relatively red \citep[e.g.,][]{zibetti2009,szomoru2013}. Luminosity-weighted ellipticities are larger, i.e., galaxies appear to be flatter. This is consistent with a simple morphological picture, where for a relatively red bulge and bluer disk, when randomly observed, the luminosity-weighting increases the ellipticity because more weight is given to the disk component. Similarly, because the velocity dispersion of the disk component is typically much lower than the bulge, a luminosity-weighted velocity dispersion is expected to be lower than a mass-weighted velocity dispersion. We find that this trend is recovered in the \ea\ simulations. Furthermore, the lower velocity dispersion will also cause the \vs\ values to be higher when using luminosity weighting as compared to mass weighting. Finally, luminosity weighted stellar ages in \ea\ are younger, consistent with observational results \citep[e.g.,][]{mcdermid2015,gonzalez2015}.

In \ha\ (Fig.~\ref{fig:lum_mass_weighting_ha}), most mock-observed quantities follow the expectations from the observations, with the exception of the size measurements. The luminosity-weighted effective radii in \ha\ are smaller than the mass-weighted quantities. We note that a burst of central star formation can make a galaxy more compact when using luminosity-weighted sizes. However, from the fact that galaxies of all sizes appear to follow the same trend, we consider this scenario to be unlikely. As the effect is small, we consider a further investigation of this result beyond the scope of the paper. We also note that while the \ha\ results for the velocity dispersion are quantitatively the same as in \ea, the difference between luminosity and mass-weighted dispersion is considerably stronger in \ha.

Lastly, all mock-observations from the \ma\ simulation follow the expected observational trends. The results are the same as for \ea: luminosity-weighted sizes are larger than the mass-weighted size, ellipticities are more flattened, velocity dispersions are considerably lower, \vse\ values are higher, and ages are younger. However, the difference in using mass-weighted versus luminosity-weighted quantities is small for all but mean stellar age.

\section{Stellar population age in SAMI, CALIFA, and \at}
\label{sec:app_age_sami_a3d}

As described in Section \ref{sec:obs_data}, mean stellar ages have been measured by \citet{scott2017} for the SAMI Galaxy Survey data, \citet{mcdermid2015} for \at\ data. Both use a method where lick indices are converted into single stellar population. For the CALIFA survey, \citet{gonzalez2015} adopt a full spectral fitting method. Besides the different methods, the wavelength regions from which the stellar population parameters have been derived are also different: 3750-5750\AA\ for SAMI, 4800-5380\AA\ for \at\, and 3745-7300\AA\ for CALIFA. Because of the different wavelength regions covered by the different surveys we expect differences in the mean age  estimates as blue wavelengths are more sensitive to young stellar populations. Therefore, in this appendix we compare the mean stellar age distribution of galaxies from the SAMI Galaxy Survey, \at, and CALIFA. 

\begin{figure*}
\includegraphics[width=0.85\linewidth]{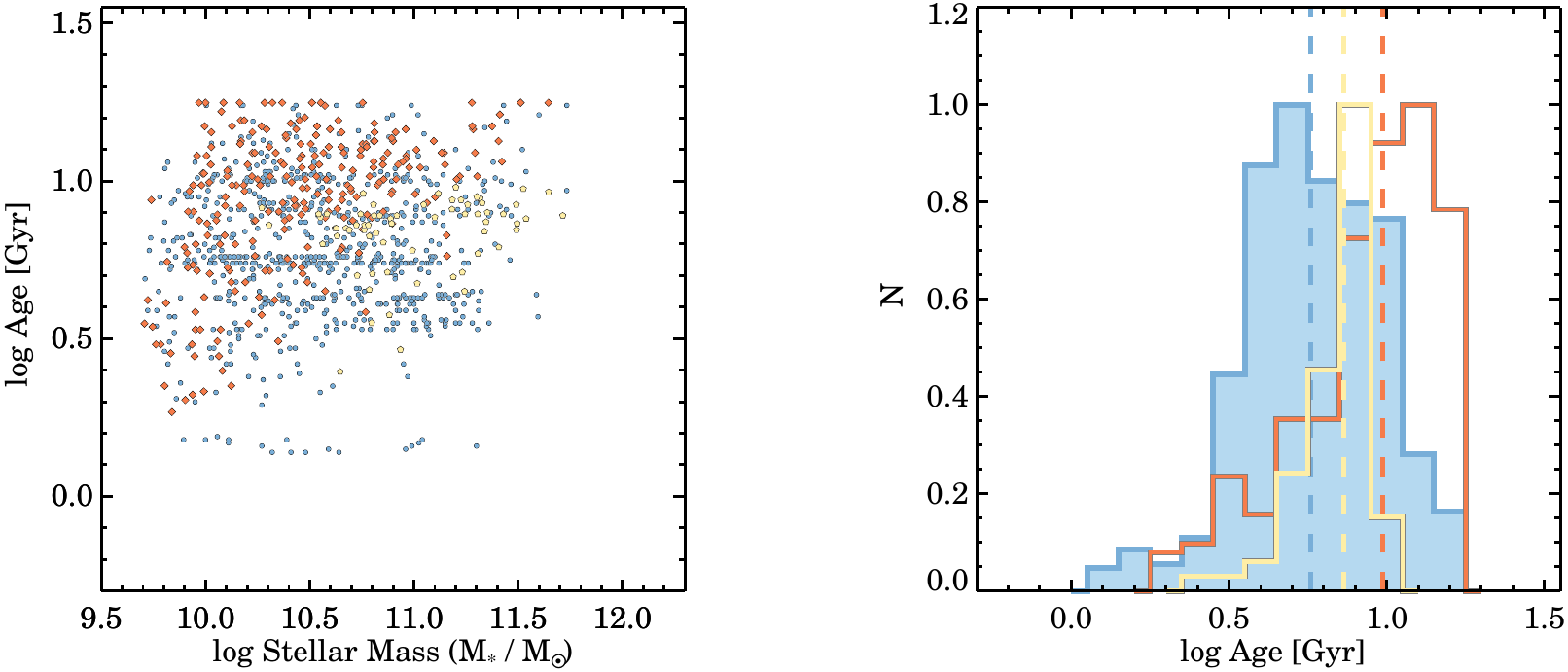}
\caption{Comparison of the mean stellar ages of early-type galaxies from the SAMI Galaxy Survey (blue), \at\ (orange), and CALIFA (yellow). At fixed stellar mass, \at\ age estimates are higher with a median $\log age$ [Gyr] of 0.99 as compared to SAMI (0.76) and CALIFA (0.87).
}
\label{fig:age_check}
\end{figure*}

In Fig.~\ref{fig:age_check} left-panel, we compare the mean $\log age$ as a function of stellar mass for the three different surveys for early-type galaxies only (E and S0). Note that the mean $\log age$ correlates significantly better with \se\ than with stellar mass \citep[e.g.,][]{mcdermid2015,scott2017}, but in order to compare with Fig.~\ref{fig:mass_age} we show stellar mass here instead of \se. We find that \at\ galaxies are clearly offset to older ages, but that this is not caused by a difference in the stellar mass distribution of the different samples. At fixed stellar mass, \at\ galaxies have higher age estimates as compared to SAMI and CALIFA. Note that there are almost no galaxies in the CALIFA Survey with ages older than 10Gyr, while there are a considerable number in SAMI and \at. This is most likely caused by the different fitting methods.

We show the distribution of $\log age$ in the right-panel of Fig.~\ref{fig:age_check}. The SAMI distribution shows the largest spread in age, but also has the lowest median $\log age$ in Gyr of 0.76. The CALIFA age distribution is considerably narrower, with a slightly higher median age (log age [Gyr] = 0.87), whereas the \at\ age distribution is noticeably offset and skewed towards older ages (median log age [Gyr] = 0.99). Thus, in order to create a homogeneous observational dataset, we subtract 0.23 dex from the \at\ ages. We do not apply an offset to the CALIFA sample because we do not find a significant offset in $\log age$ for the entire SAMI and CALIFA sample without a morphological cut (median $\log age$ in Gyr of 0.63 and 0.61, respectively).

\bsp	
\label{lastpage}
\end{document}